\documentclass[twocolumn]{aa}
\usepackage{natbib}
\bibpunct{(}{)}{;}{a}{}{,} 
\usepackage{graphicx}
\usepackage{txfonts}

\usepackage{xspace}
\usepackage{supertabular}
\usepackage{longtable}
\usepackage[version=3]{mhchem}
\usepackage[OT2,OT1]{fontenc}
\newcommand\cyr
{
	\renewcommand\rmdefault{wncyr}
	\renewcommand\sfdefault{wncyss}
	\renewcommand\encodingdefault{OT2}
	\normalfont
	\selectfont
}
\DeclareTextFontCommand{\textcyr}{\cyr}

\newcommand{\FastChem}{\texttt{FastChem}\xspace}
\newcommand{\kms}{km~s$^{-1}$\xspace}
\newcommand{\continuumpressure}{3.5 mbar\xspace}
\newcommand{\maxfeiiwindspeed}{$0.18 \pm 0.27$ \xspace}
\newcommand{\maxfeiwindspeed}{$0.84 \pm 0.37$ \xspace}
\newcommand{\stellarmass}{1.978 \pm 0.023 M_{\odot}\xspace}
\newcommand{\stellarradius}{2.178 \pm 0.011 R_{\odot}\xspace}
\newcommand{\planetmass}{2.44 \pm 0.70 M_{J}\xspace}
\newcommand{\planetradius}{1.783 \pm 0.009 R_{J}\xspace}
\newcommand{\orbitalvelocity}{234.24 $\pm$ 0.90\xspace}
\newcommand{\gcm}{g~cm$^{-3}$\xspace}

\begin{document}

   \title{A spectral survey of an ultra-hot Jupiter}

   \subtitle{Detection of metals in the transmission spectrum of KELT-9 b}

   \author{H.~J. Hoeijmakers\inst{1,2}
          \and
          D. Ehrenreich\inst{1}
          \and
          D. Kitzmann\inst{2}
          \and
          R. Allart\inst{1}
          \and
          S.~L. Grimm\inst{2}
          \and
          J.~V. Seidel\inst{1}
          \and
          A. Wyttenbach\inst{3}
          \and
          L. Pino\inst{4}
          \and
          L.~D. Nielsen\inst{1}
          \and
          C. Fisher\inst{2}
          \and
          P.~B. Rimmer\inst{5,6}
          \and
          V. Bourrier\inst{1}
          \and
          H.~M. Cegla\inst{1}\thanks{CHEOPS Fellow}
          \and
          B. Lavie\inst{1}
          \and
          C. Lovis\inst{1}
          \and
          A.~B.~C. Patzer\inst{7}
          \and
          J.~W. Stock\inst{8}
          \and
          F.~A. Pepe\inst{1}
          \and
          Kevin Heng\inst{2}
   }

   \institute{Observatoire de Gen\`eve, University of Geneva,
              Chemin des Maillettes, 1290 Sauverny, Switzerland\\
              \email{jens.hoeijmakers@unige.ch}
         \and
             Center for Space and Habitability, Universit\"at Bern, Gesellschaftsstrasse 6, 3012 Bern, Switzerland\\
             \email{jens.hoeijmakers@space.unibe.ch}
         \and
             Leiden Observatory, Universiteit Leiden, Niels Bohrweg 2, 2333 CA Leiden, The Netherlands
             \and
             Anton Pannekoek Institute of Astronomy, University of Amsterdam, Science Park 904, 1098 XH Amsterdam, The Netherlands
   \and
             Cavendish Astrophysics Battcock Centre for Experimental Astrophysics, Cavendish Laboratory, Cambridge University, JJ Thomson Avenue, CB3 0HE Cambridge, United Kingdom
         \and
             MRC Laboratory of Molecular Biology, Francis Crick Avenue, Cambridge Biomedical Campus, Cambridge University, CB2 0QH Cambridge, United Kingdom
         \and
             Zentrum für Astronomie und Astrophysik, Technische Universität Berlin, Hardenbergstrasse 36, D-10623 Berlin,
Germany
         \and
             Department of Chemistry and Environmental Science, Medgar Evers College, City University of New York, 1650 Bedford
Avenue, Brooklyn, NY 11235, USA}

   \date{Received January 18, 2019; accepted May 3, 2019}


  \abstract
   {KELT-9 b exemplifies a newly emerging class of short-period gaseous exoplanets that tend to orbit hot, early type stars - termed \textit{ultra-hot Jupiters}. The severe stellar irradiation heats their atmospheres to temperatures of $\sim 4,000$ K, similar to the photospheres of dwarf stars. Due to the absence of aerosols and complex molecular chemistry at such temperatures, these planets offer the potential of detailed chemical characterization through transit and day-side spectroscopy. Detailed studies of their chemical inventories may provide crucial constraints on their formation process and evolution history.}
   {To search the optical transmission spectrum of KELT-9 b for absorption lines by metals using the cross-correlation technique.}
   {We analyze two transit observations obtained with the HARPS-N spectrograph. We use an isothermal equilibrium chemistry model to predict the transmission spectrum for each of the neutral and singly-ionized atoms with atomic numbers between 3 and 78. Of these, we identify the elements that are expected to have spectral lines in the visible wavelength range and use those as cross-correlation templates.}
   {We detect $(>5 \sigma)$ absorption by \ion{Na}{i}, \ion{Cr}{ii}, \ion{Sc}{ii} and \ion{Y}{ii}, and confirm previous detections of \ion{Mg}{i}, \ion{Fe}{i}, \ion{Fe}{ii} and \ion{Ti}{ii}. In addition, we find evidence of \ion{Ca}{i}, \ion{Cr}{i}, \ion{Co}{i}, and \ion{Sr}{ii} that will require further observations to verify. The detected absorption lines are significantly deeper than predicted by our model, suggesting that the material is transported to higher altitudes where the density is enhanced compared to a hydrostatic profile, i.e. that the material is part of an extended or outflowing envelope. There appears to be no significant blue-shift of the absorption spectrum due to a net day-to-night side wind. In particular, the strong \ion{Fe}{ii} feature is shifted by \maxfeiiwindspeed \kms, consistent with zero. Using the orbital velocity of the planet we derive revised masses and radii of $M_* = \stellarmass$, $R_* = \stellarradius$, $m_p = \planetmass$ and $R_p = \planetradius$.}
   {}


   \maketitle
%

   \section{Introduction}
   Hot Jupiters provide a unique window into the nature of the exoplanet population due to the combination of their size, high temperature and short orbital periodicity. Furthermore, their atmospheres are often inflated \citep{Baraffe2010}, making them especially amenable for transmission spectroscopy during transit events. During a transit some starlight passes through the upper atmosphere of the planet, where certain wavelengths are absorbed depending on its chemical composition. This wavelength-dependent absorption manifests itself as a wavelength-dependence of the apparent radius of the planet. The spectrum of the atmosphere generally has two components: line-absorption caused by discrete energy transitions in atoms and molecules, and continuum absorption or scattering caused by molecules or aerosol particles.

   The identification and measurement of the relative strengths of spectral lines and bands in the transmission spectra of exoplanets have been used to put constraints on the atmospheric composition and thermal structure \citep[see e.g.][]{VidalMajar2003, Redfield2008, Sing2008, Huitson2012, Wyttenbach2015, Nikolov2018, Spake2018}, with the ultimate aim to provide constraints on their formation scenario and evolution \citep{Madhusudhan2014, Kreidberg2015, Mordasini2016, Brewer2017}. In addition, these lines may be used to probe atmospheric dynamics because velocity streams in the atmosphere cause the spectral lines to be Doppler-shifted away from the rest-frame velocity of the planet \citep{Snellen2010, Kempton2012, Showman2013, Kempton2014, Louden2015, Brogi2016,Allart2018}.

   Access to the information contained in individual absorption lines requires high spectral resolution in the order of $R=\frac{\lambda}{\Delta\lambda}\sim100,000$. Such spectral resolution is generally limited to ground-based instrumentation, mostly in the form of optical and NIR echelle spectrographs. When seen in transmission, the spectral lines formed in the lower part of the atmosphere of a typical hot Jupiter can have depths of up to $\sim10^{-3}$ times the flux of the host star. To reach this sensitivity with high-resolution spectrographs on present-day telescopes, many targeted absorption lines can be combined using a form of cross-correlation. Besides raising the sensitivity by averaging out the photon noise, the use of cross-correlation has the advantage of providing additional robustness against spurious spectral features, because the distribution of absorption lines is unique for each species, and follows the radial velocity of the planet \citep{Snellen2010, Brogi2012}. As such, enhancements of the cross-correlation function are hard to mimic by systematic noise that is not correlated with the absorption line spectrum of the target species.

   A theoretical study by \citet{Kitzmann2018} predicted the presence of absorption lines of iron (Fe) in the transmission spectrum of the ultra-hot Jupiter KELT-9 b. KELT-9 b orbits the bright A0 star HD 195689 with a periodicity of 1.48 days \citep{Gaudi2017}. Due to its close orbital distance of approximately 0.03 AU \citep{Gaudi2017}, it is heated to an equilibrium temperature of $4050 \pm 180$ ~K, similar to the photospheres of many dwarf stars. Such temperatures are high enough to justify simplifying assumptions of equilibrium chemistry and a nearly purely atomic gas at the dayside and terminator regions \citep{Arcangeli2018,Kitzmann2018, Lothringer2018, Parmentier2018}. This potentially greatly simplifies the interpretation of the transmission spectrum of KELT-9 b when compared to other hot Jupiters for which aerosols and non-equilibrium chemistry are important factors. However, the night-side of the planet is expected to be in a much cooler regime that is likely characterized by complex molecular chemistry and condensation. Condensed droplets and grains may rain out to the interior of the planet, potentially depleting these species from the atmosphere in what is known as a cold-trap \citep{Spiegel2009}. Depending on the circulation efficiency, these species may then be transported back to the hot day side where they can again evaporate into the gas phase and dissociate to produce atomic absorption lines that can be observed in the transmission spectrum \citep{Showman2009, Fortney2010}. The observed transmission spectrum is therefore expected to be linked to the chemistry elsewhere on the planet via the global circulation pattern. Regardless, chemical equilibrium is a valid assumption on the hot day-side and terminator region due to the efficiency of thermal reactions at these high temperatures, given the elementary abundances at these locations as set by global circulation processes \citep{Kitzmann2018}. 

   Seeking to validate the predictions by \citet{Kitzmann2018} we previously applied the cross-correlation technique to transit observations obtained with the high-resolution HARPS-North (HARPS-N) spectograph and found the presence of absorption lines by neutral iron (\ion{Fe}{i}), but also strong lines of ionized iron (\ion{Fe}{II}) and titanium (\ion{Ti}{ii}) \citep{Hoeijmakers2018a}. In addition, Balmer lines of hydrogen where observed by \citet{Yan2018} and \citet{Cauley2018} in the transmission spectrum of the same planet, as well as individual lines of \ion{Fe}{i}, \ion{Fe}{ii} and the \ion{Mg}{i} triplet \citep{Cauley2018}.

   In July of 2018 we obtained a second transit observation with the HARPS-N spectrograph and proceeded to perform a survey for additional species that have absorption lines in the optical range of HARPS-N, using the high-resolution cross-correlation technique. This survey includes all neutral and singly ionized species with atomic numbers 3 to 78 (lithium to platinum)\footnote{We chose to limit the survey to elements lighter than platinum because spectral data becomes sparse and elemental abundances diminish for elements with higher atomic numbers. Helium is not expected to feature strong lines in the optical, and the strong hydrogen lines will be treated in a separate study.} for which adequate line-list data was available. This paper proceeds to describe the observations and the analysis method in Section \ref{sec:obs} (with a detailed description of our application of the cross-correlation operation in Appendix \ref{app:ccv}), the cross-correlation templates in Section \ref{sec:templates} and the results of the survey in Section \ref{sec:results}.

\section{Observations and cross-correlation analysis}\label{sec:obs}
We have observed two transits with the HARPS-North spectrograph mounted on the 3.58-meter Telescopio Nazionale Galileo (TNG) in La Palma, Spain, on 31-07-2017 (programme A35DDT4, hereafter Night 1) and on 20-07-2018 (programme OPT18A-38, hereafter Night 2). Both programmes have been carried out in the frame work of the SPADES survey (``Sensing Planetary Atmospheres with Differential Echelle Spectroscopy''). Results from Night~1 have been previously published in \citet{Hoeijmakers2018a}.

\begin{figure*}
  \centering
  \includegraphics[width=18cm]{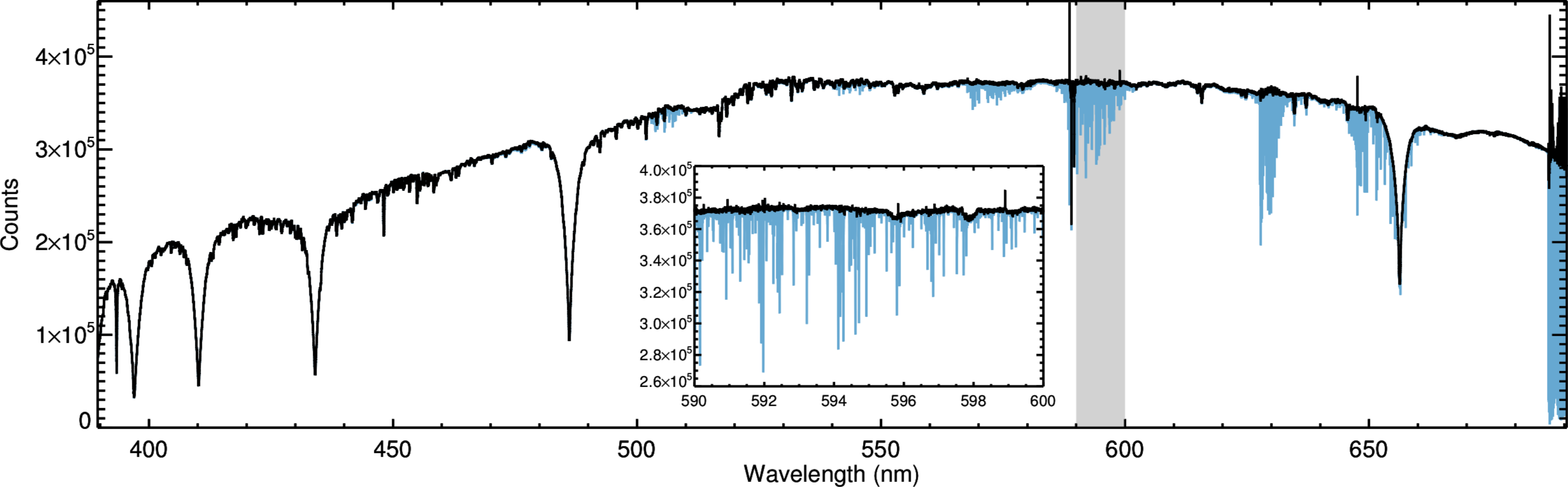}
  \caption{The mean out-of-transit spectrum of the KELT-9 system as observed by HARPS-N, produced by stitching together the individual spectral orders after correction of the blaze function. The blue and black lines show the spectrum before and after telluric correction. The smaller panel shows the telluric water band that is highlighted in gray in detail, demonstrating the effect of the telluric correction. The outlying points are residuals of deep telluric absorption and a small number of emission lines that have remained uncorrected.}%
  \label{fig:spectrum}
\end{figure*}

HARPS-North is an optical spectrograph that covers wavelengths between 387.4 and 690.9 nm at a spectral resolution of $R\sim 115,000$, corresponding to a Doppler velocity of $\sim 2.7$ \kms. The observations were obtained during both 3.9-hour transits, as well as during 5.0 and 4.5 hours baseline before and after the transit events. The exposure time was fixed to 600 s, yielding runs of 49 and 46 exposures, 19 of which took place during either transit (see Table \ref{tab:obs}). The observations of the second night were affected by a loss of flux in the bluest orders caused by an error in the control software of the Atmospheric Dispersion Corrector (ADC). Fig. \ref{fig:spectrum} shows the average one-dimensional spectrum of the KELT-9 system as observed by HARPS-N.

\begin{table}
\caption{Overview of the observations with HARPS-N.}             
\label{tab:obs}      
\centering                          
\begin{tabular}{l c c}        
\hline\hline                 
               & Night 1 & Night 2 \\    
\hline                        
Date           & 31-07-2017 & 20-07-2018     \\      
$t_{\textrm{start}}$ (UT)  & 20:59  & 21:20   \\
$T_{c}$ (UT)      & 01:59 & 01:43 \\
$t_{\textrm{end}}$ (UT)    & 05:19 &  05:09 \\
$N_{\textrm{exp}}$         & 49    & 46 \\
$t_{\textrm{exp}}$ (s)     & 600   & 600 \\
\hline                                   
\end{tabular}
\end{table}

The raw spectra were reduced using the HARPS Data Reduction Software (DRS, version 3.8) which interpolates the extracted spectra on a uniform grid with a spacing of 0.01\AA. Absorption lines of the Earth atmosphere (telluric contamination) were corrected for using the MOLECFIT package \citep{Smette2015} following \citet{Allart2017}, and the subsequent analysis followed the general procedure from previous studies that use the high-resolution cross-correlation technique \citep[e.g.][]{Snellen2010, Brogi2012}, in particular the analyses used in \citet{Hoeijmakers2018a} for the data of Night 1, and \citet{Hoeijmakers2018b}.

For reasons of computational efficiency, we subdivided the one-dimensional spectra in 15 sections of 20 nm wide (each containing 20,000 spectral pixels)\footnote{The HARPS range is divided into 15 bins of 20 nm each, with the reddest few nm being ignored due to the presence of the strong oxygen band near 687 nm.}, and performed the following analysis on each section individually. To isolate the signature of the planet, each spectrum was normalized to its median value and divided through the mean of the out-of-transit spectra, $F(\lambda,t_{\textrm{out}})$. We then constructed an outlier-mask by flattening each spectrum using a median filter (to remove broadband continuum variations) followed by applying a running standard deviation, both with a width (in wavelength) of 75 pixels. All values that were more than $5\sigma$ away were flagged as NaNs and the previous normalization and removal of the mean out-of-transit spectrum were repeated. The resulting residuals were then filtered using a Gaussian high-pass filter with a $1\sigma$ width of 75 pixels, and any remaining $5\sigma$ outliers were again flagged and added to the outlier mask. This amounted to a total of 1729 and 8712 spectral pixels in nights 1 and 2 respectively. In addition, we masked out selected regions with systematic residuals inside strong telluric absorption lines (notably residuals caused by the O$_2$ bands) and regions with low signal-to-noise. In this way 23,226 and 72,542 pixels were removed additionally. In total, $0.38\%$ of all spectral pixels were masked out.

This procedure effectively yields the transmission spectrum $x(\lambda,t)=\frac{F(\lambda,t)}{F(\lambda,t_{\textrm{out}})}-1$, as the ratio between each spectrum obtained at time $t$ divided by the average out-of-transit spectrum.

\smallskip

The cross-correlation, $c$, is performed using a weighted mask (template), $T$, that acts to co-add the individual spectral pixels in the transmission spectrum observed at time, $t$, for velocity shifts between $\pm 1000$ \kms in steps of 2 \kms. This procedure follows the practise of \citet{Baranne1996}, \citet{Pepe2002}, \citet{Allart2017} a.o. and is prescribed in Eq. \ref{eq:ccv}:

\begin{equation}\label{eq:ccv}
c(v,t) = \sum_{i=0}^{N}x_i(t) T_i(v)
\end{equation}

Here $x_i(t)$ are each of the $N$ spectral points in the transmission spectrum obtained at time $t$, $T_i(v)$ are the corresponding values of the template that is Doppler shifted to a radial velocity, $v$. $T$ takes on non-zero values inside a spectral line of interest, and zero in the continuum. Furthermore, $T$ is normalized such that $\sum_{i=0}^{N}T_i(v) = 1$. This operation thus effectively computes a weighted average of the absorption in the transmission spectrum at the location of the spectral lines included in the template. Because $c(v,t)$ is computed for each 20 nm bin and each exposure of the time-series, we co-add these into a one-dimensional measurement of the average line profile in-transit. This average is weighted for the signal-to-noise ratio obtained in each exposure, in each spectral bin, as detailed in Appendix \ref{app:ccv}. This procedure is consistent with the approach applied in \citet{Hoeijmakers2018a}.

\begin{figure*}
   \centering
   \includegraphics[width=19cm]{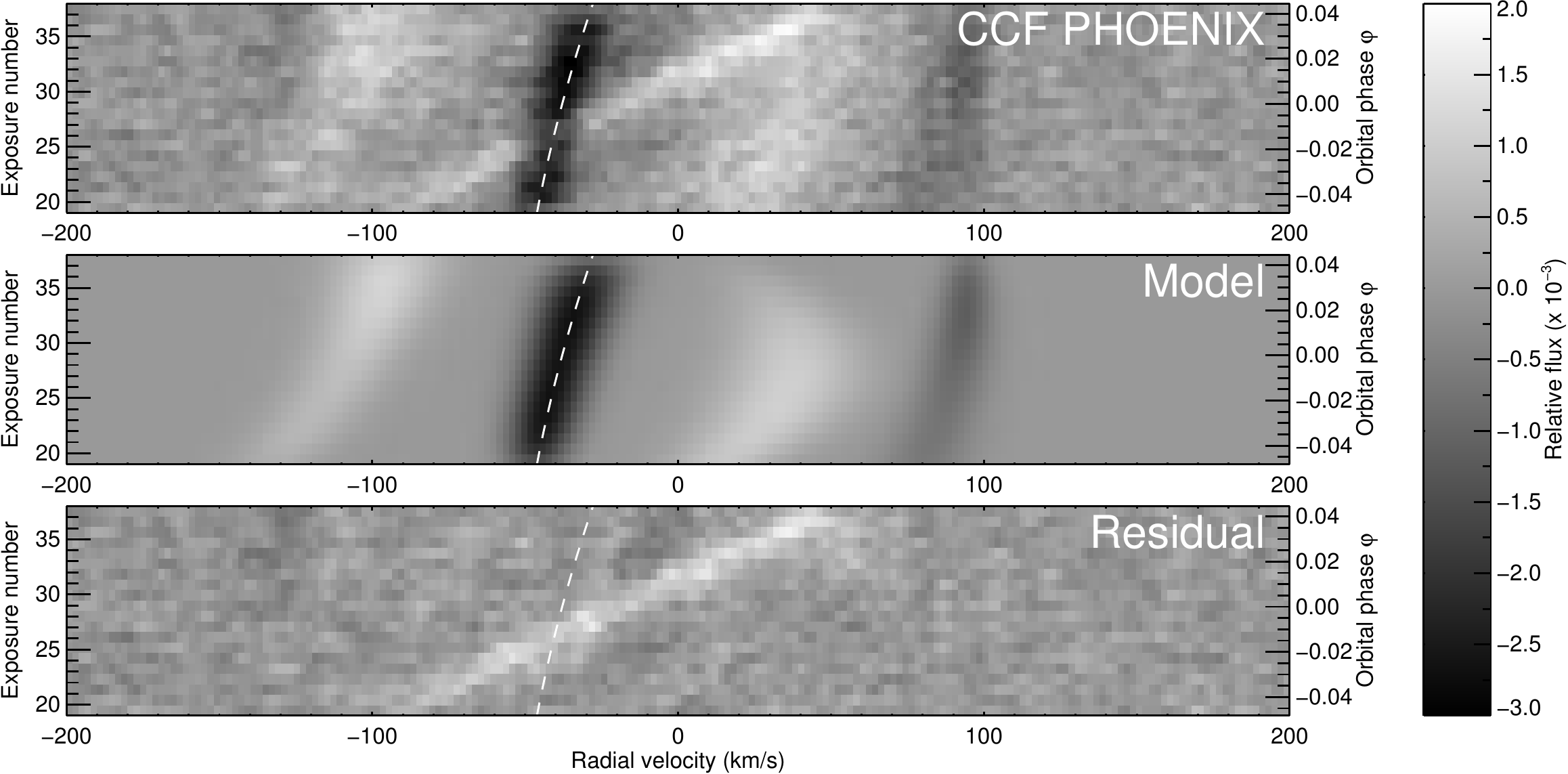}
      \caption{Removal of the Doppler shadow for the first night of data. The top panel shows the two-dimensional cross-correlation function as obtained when using a PHOENIX 10,000 K stellar atmosphere model template. The dark feature near $-40$ km/s is the Doppler shadow caused by the planet obscuring part of the disk-integrated stellar line, that is broadened by $v \sin i = 111.4$ \kms \citep{Gaudi2017}. The bright linear slanted feature is the signature of the atmosphere of the planet. It appears in this cross-correlation because the star and the planet have ionized lines in common \citep{Hoeijmakers2018a}. It is ignored when constructing and fitting the model of the Doppler shadow. Besides these, there are significant cross-correlation residuals due to stellar line-shape variations that we tentatively attribute to pulsations of KELT-9 that are not characterized in the literature as of yet, but seem to resemble those that have been observed in hot $\delta$ Scuti stars such as $\beta$ Pictoris \citep{Koen2003} and Wasp-33 \citep{Colliercameron2010}. The middle panel shows the full model as fit to this cross-correlation function, where the dashed line indicates the polynomial fit to the centroid velocities of the main shadow feature in each exposure. The bottom panel shows the residuals after removal.}
         \label{fig:dopplershadow}
   \end{figure*}

We follow \citet{Hoeijmakers2018a} in producing an empirical model of the Doppler shadow that is caused by the obscuration of only part of the stellar absorption lines, effectively deforming them during the planetary transit (see Fig. \ref{fig:dopplershadow}). We characterize the shadow by cross-correlating the data with a continuum-normalized PHOENIX model of the stellar photosphere corresponding to an effective temperature of 10,000 K, log$(g) = 4.5$ and solar metallicity \citep{Husser2013}. The resulting cross-correlation profile is fit independently in each exposure using a Gaussian model. The resulting two-dimensional profile of $c(v,t)$ is then modeled by fitting low-order polynomials through the obtained Gaussian fit parameters (center position, amplitude and width), essentially requiring that the fit parameters of the Doppler shadow vary smoothly in time. Initially, this is only done for those exposures in which the signature of the Doppler shadow does not overlap with the signature of the atmosphere of the planet (the velocity of these overlaps during part of the transit). Then this model is subtracted, allowing the residual signature of the planet to be modelled. The initial Doppler-shadow is fit again after subtracting the signature of the planet, this time including the region where the radial velocity of the planet and the Doppler shadow overlap. The same procedure is done for four additional signatures that may be associated with stellar pulsations, though these mostly occur at velocities that do not coincide with the velocity of the planet at any time during the transit (see Fig \ref{fig:dopplershadow}).

The resulting two-dimensional model of the Doppler-shadow is subtracted from the cross-correlation profiles obtained for each of the various cross-correlation templates, multiplied by a scaling factor that minimizes the sum of the squared residuals - but ignoring cross-correlation values at which the planet signature is expected to be present. The values of these scaling factors are provided in Table \ref{table:DSfitting}.

\section{Cross-correlation templates}\label{sec:templates}
To compute the cross-correlation function, a mask is needed that defines the weights by which spectral pixels are co-added as per Eq. \ref{eq:ccv}. We construct a mask for each species by forward modeling the transmission spectrum of KELT-9 b, assuming that it contains line-absorption of each species individually.

\subsection{Opacity functions}
Line-opacity functions are computed using HELIOS-K \citep{Grimm2015}, adopting the transition tables from \citet{Kurucz2018}. As the partition functions in the database by Kurucz are not complete up to high atomic number, the partition functions were instead computed using the energy levels and statistical weights provided by the NIST database when needed \citep{Kramida2018}. Continuum opacity functions due to collision-induced absorption of H-He, H$_2$-H$_2$ and H$_2$-He were adopted from HITRAN \citep{Richard2012}, and  H$^-$ bound-free and free-free absorption from \citet{John1988}. The \citet{Kurucz2018} database includes broadening by radiation dampening, but not pressure broadening. This limitation does not significantly affect the analysis because the high-resolution transmission spectrum probes low pressures and the pressure-broadened wings are masked by the H$^{-}$ continuum which occurs at a pressure of around \continuumpressure in these simulations \citep{Kitzmann2018}.

\subsection{Chemistry}
Chemistry calculations are done with the open source code \texttt{FastChem} \citep{Stock2018}. \FastChem calculates the chemical equilibrium composition (i.e. assuming that all reaction rates are dominated by thermal timescales), and natively includes 560 chemical species. In addition, we add approximately 180 new ions to the set of species, i.e. singly and doubly ionized atoms as well as some anions, for most elements lighter than neptunium (Z=93). Because many of the required equilibrium constants are not available in the usual chemical databases \citep[e.g.][]{Chase1998nist}, we calculate them using the Saha equation when needed \citep{Saha1920}.

For the ionisation reaction of, for example, caesium,
\begin{equation}
\ce{Cs+ + e- -> Cs} \ ,
\end{equation}
the corresponding temperature-dependent equilibrium constant for \ce{Cs+} is given by
\begin{equation}\label{eq:eqconst}
K(T) = \frac{n_e n_{\ce{Cs+}} }{ n_{\ce{Cs}} } \ ,
\end{equation}
where $n$ are the corresponding number densities of \ce{Cs}, \ce{Cs+}, and the free electrons, respectively.

To compute the right hand side of Eq. \ref{eq:eqconst} we employ the Saha equation, that describes the relation between the number densities $n$ of two ionisation levels $i$ and $i+1$ at a given temperature,
\begin{equation}\label{eq:saha}
  \frac{n_e n_{i+1}}{n_i} = \frac{2}{\lambda^3} \frac{Q_{i+1}}{Q_i}\exp \left( - \frac{\Delta E_{i+1,i}}{k_B T} \right) \ ,
\end{equation}
where $\lambda$ is the thermal de Broglie wavelength of an electron, $Q$ the corresponding partition function, and $\Delta E$ the energy difference between the two ionisation levels. The partition functions are calculated based on the line transition database by \citet{Kurucz1995all..book.....K} where available, and from those in the NIST Atomic Spectra Database \citep{NIST_ASD}, otherwise. The ionisation energies are taken from the \citet{CRC_handbook85Ed}.

For doubly-ionised species, the Saha equation is applied twice: Once to evaluate the ratio
\begin{equation}
  \frac{n_{\ce{Cs+}}}{n_e n_{\ce{Cs++}} } \ ,
\end{equation}
and once more for the corresponding equation (Eq. \ref{eq:eqconst}) for the number density of the singly-ionised species. In the case of anions, $\Delta E$ in the Saha equation is replaced by the electron affinity.

The resulting mass action constants are fitted as a function of temperature according to the equation used within \texttt{FastChem}
\begin{equation}
\ln \overline{K}_i(T) = \frac{a_0}{T} + a_1 \ln T + b_0 + b_1 T + b_2 T^2 \ ,
\end{equation}
where $\overline{K}$ is the mass action constant with respect to a reference pressure of 1 bar (see \citet{Stock2018} for details). The resulting calculated coefficients for temperatures between 100 K and 6000 K can be found in Table \ref{table:FastChemData}. Solar elemental abundances are taken from \citet{Asplund2009ARA&A..47..481A}, either based on their solar photospheric value where available, or from their meteoritic abundance otherwise.

\begin{figure}
   \centering
   \includegraphics[width=7cm,angle=270]{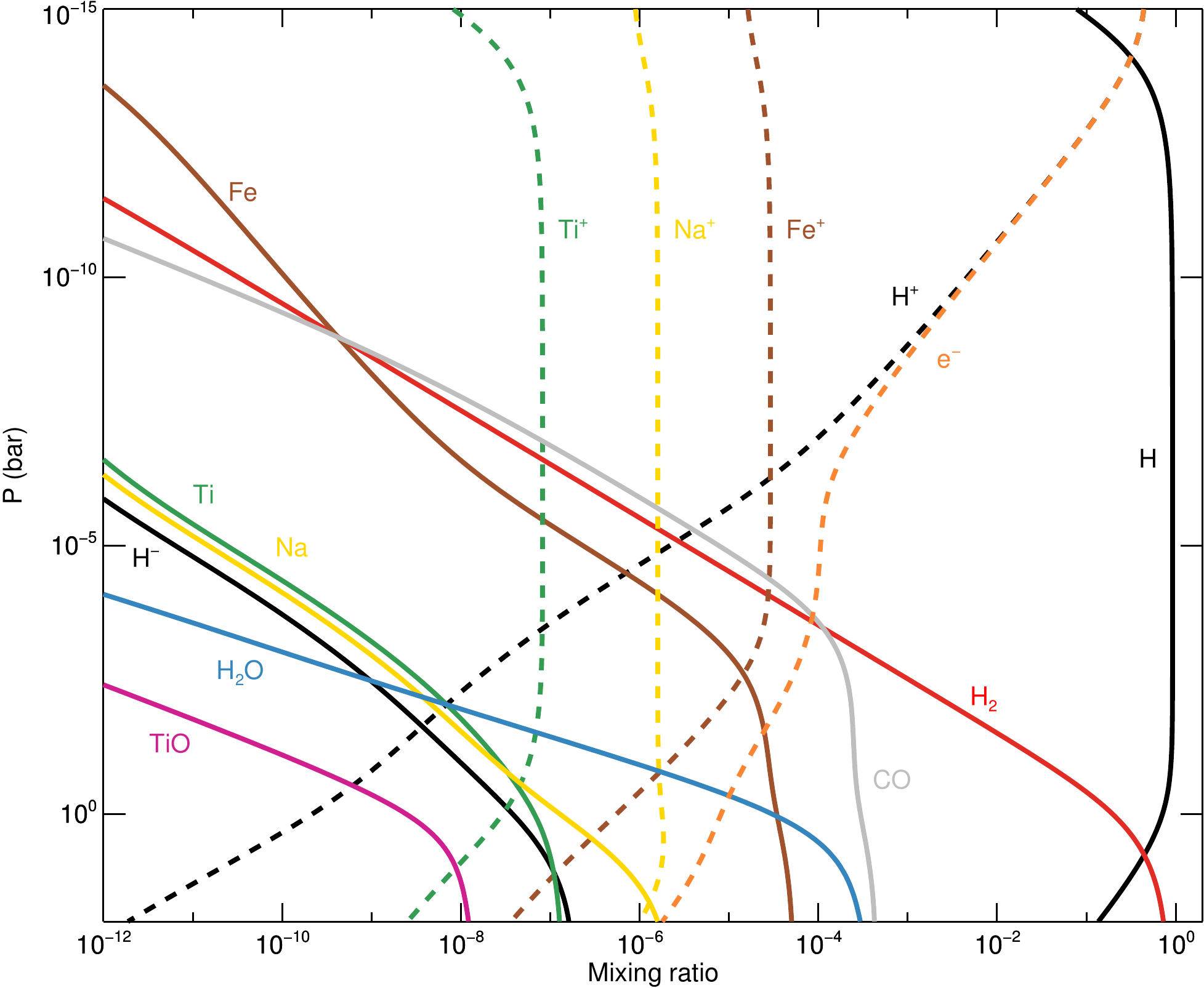}
   \caption{Abundance profiles modeled with \texttt{FastChem} for a selection of neutral atoms (solid lines) and ions (dashed lines), as well as molecules, H$^-$ and electrons. The model assumes an isothermal structure at 4,000 K and solar metallicity. Over most of the modeled pressure region, molecules tend to be dissociated and metals tend to be singly ionized. The decrease of ion abundances towards higher altitudes (pressures below $\sim10^{-12}$ bars) indicates the emergence of doubly ionized species at low pressures. The effects of photo-ionization are not included, as these are predicted to be small in the denser parts of the atmosphere \citep{Kitzmann2018}.}%
   \label{fig:abundances}
   \end{figure}

To calculate the cross-correlation templates, we assume elemental abundances equal to solar metallicity and an isothermal profile of 4,000 K between pressures of 10 bar to $10^{-15}$ bar. At this temperature, the abundances of molecules are low and atoms are partially ionized (see Fig. \ref{fig:abundances}). Significant abundances of doubly ionized species only occur at pressures below about $10^{-12}$ bars (see Fig. \ref{fig:abundances}), where the density and optical depth are small. In this model, the abundances of the neutral elements plus their first ionized state therefore approximately correspond to the assumed metallicity value, consistent with \citet{Kitzmann2018}, and we consider only neutral and singly ionized species in this analysis.

\subsection{Model transmission spectra}\label{sec:transmission_spectra}
\begin{figure*}
  \centering
  \includegraphics[width=18cm]{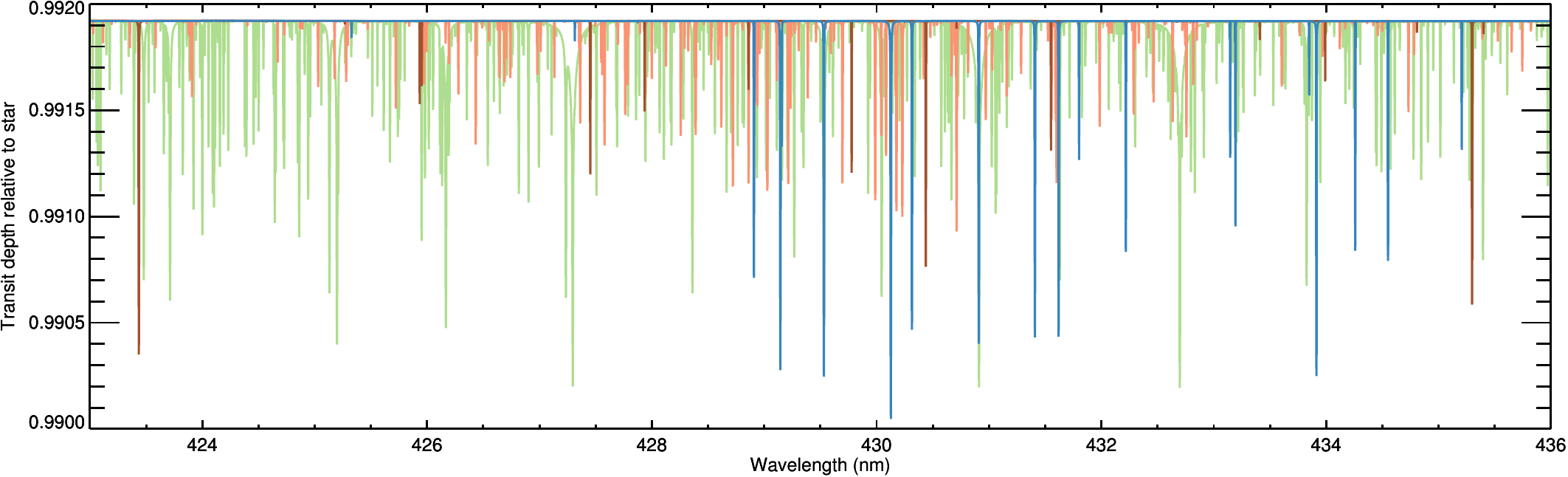}
  \caption{Sample of model spectra of \ion{Fe}{I}, \ion{Fe}{II}, \ion{Ti}{I} and \ion{Ti}{II} in green, brown, red and blue respectively, over a 13 nm slice of the HARPS-N spectral range. These spectra are calculated assuming an isothermal atmosphere at 4,000 K between pressures of 10 to $10^{-15}$ bars and abundances correponding to the solar metallicity value (Section \ref{sec:transmission_spectra}). A full overview of all the calculated model spectra is provided in Fig. \ref{fig:spectra}. Together, these figures serve to show the rich absorption spectra caused by the presence of these atoms in the atmosphere of KELT-9 b.}
  \label{fig:specsample}
\end{figure*}

The model transmission spectra are simulated using a custom-built raytracing algorithm following the method described in \citet{Gaidos2017}. We assumed a variable-gravity, isothermal atmosphere at a temperature of 4,000 K between pressures of 10 to $10^{-15}$ bars, and single-elemental abundances correponding to the solar metallicity value for each of the species included in this survey, along with continuum absorption by H$^-$ and scattering by H and H$_2$. A sample of the thus computed spectra for \ion{Fe}{I}, \ion{Fe}{II}, \ion{Ti}{I} and \ion{Ti}{II} is shown in Fig. \ref{fig:specsample}, and Fig. \ref{fig:spectra} shows the computed transmission spectra of the full survey. Application of the cross-correlation function is limited to species that have multiple absorption lines protruding through the continuum in the observed waveband. Of the species between atomic numbers 3 and 78, we selected \ion{Na}{i}, \ion{Mg}{i}, \ion{Si}{i}, \ion{Ca}{i}, \ion{Sc}{I} \ion{Sc}{ii}, \ion{Ti}{i}, \ion{Ti}{ii}, \ion{V}{i}, \ion{V}{ii}, \ion{Cr}{i}, \ion{Cr}{ii}, \ion{Mn}{i}, \ion{Mn}{ii}, \ion{Fe}{i}, \ion{Fe}{ii}, \ion{Co}{i}, \ion{Ni}{i}, \ion{Sr}{ii}, \ion{Y}{i}, \ion{Y}{ii}, \ion{Zr}{i}, \ion{Zr}{ii}, \ion{Nb}{i}, \ion{Nb}{ii}, \ion{Ru}{i}, \ion{Ba}{ii}, \ion{La}{ii}, \ion{Ce}{ii}, \ion{Pr}{ii}, \ion{Nd}{ii}, \ion{Sm}{ii}, \ion{Eu}{ii}, \ion{Gd}{ii}, \ion{Tb}{ii}, \ion{Dy}{ii}, \ion{Er}{ii}, \ion{W}{i} and \ion{Os}{i} as candidates with significant line absorption in the HARPS-N waveband. These are highlighted in Fig. \ref{fig:spectra}.

The model spectra of these species are convolved with a Gaussian kernel with a FWHM width of $0.8$ \kms to match the approximate wavelength sampling of HARPS-N. They are then converted to cross-correlation templates by fitting and subtracting the continuum with a third-order polynomial and applying a threshold filter that sets any value smaller than 0.5\% of the maximum absorption line to zero. Finally, they are interpolated onto the wavelength grid of the data ($x_i$), multiplied by a Doppler-factor $1+v/c$ to shift over a range of radial velocities during cross-correlation, and normalized to unity. Through the CDS  \citep{Genova2000} we make a subset of these cross-correlation templates publicly available (see Section \ref{sec:results}).

\subsection{Model injection}\label{sec:injection}
Like in previous works \citep{Hoeijmakers2015,Hoeijmakers2018b}, the simulated transmission spectra are also used as forward-models by injecting them into the pipeline-reduced data using the known system parameters as published by \citet{Gaudi2017}, and processing this contaminated data in the same way as the data without injected model (i.e. the ``un-contaminated'' data). The retrieved cross-correlation strengths of the injected spectra inform the weights of the co-addition of the cross-correlation values for each wavelength bin and exposure (see Appendix \ref{app:ccv}). In addition, this injection of the model templates provides measurements of the average spectral lines as predicted by the models of \citet{Kitzmann2018}, against which any detected signals can be compared. Prior to injection, the model templates are rotation-broadened assuming a rotation period consistent with the orbital period of the planet (i.e. assuming tidal locking). The rotation profile simultaneously models rigid-body rotation and the spectral resolution of the HARPS-N instrument, following the formalism presented by \citet{Brogi2018}. In addition, we perform a box-smooth\footnote{A box-smooth or top-hat filter in velocity space is equivalent to convolution with a normalized kernel that has a non-zero value for velocities smaller than some limiting velocity, i.e. $|v| < v_{\textrm{limit}}$, and is equal to zero elsewhere.} to account for the changing radial velocity of the planet during a single 600 s exposure. After correlation, the contaminated and un-contaminated cross-correlation functions obtained in this way are subtracted, yielding the excess absorption due to the injected model.

\section{Results and discussion}\label{sec:results}

\begin{figure*}
   \centering
   \includegraphics[height=19cm,angle=-90]{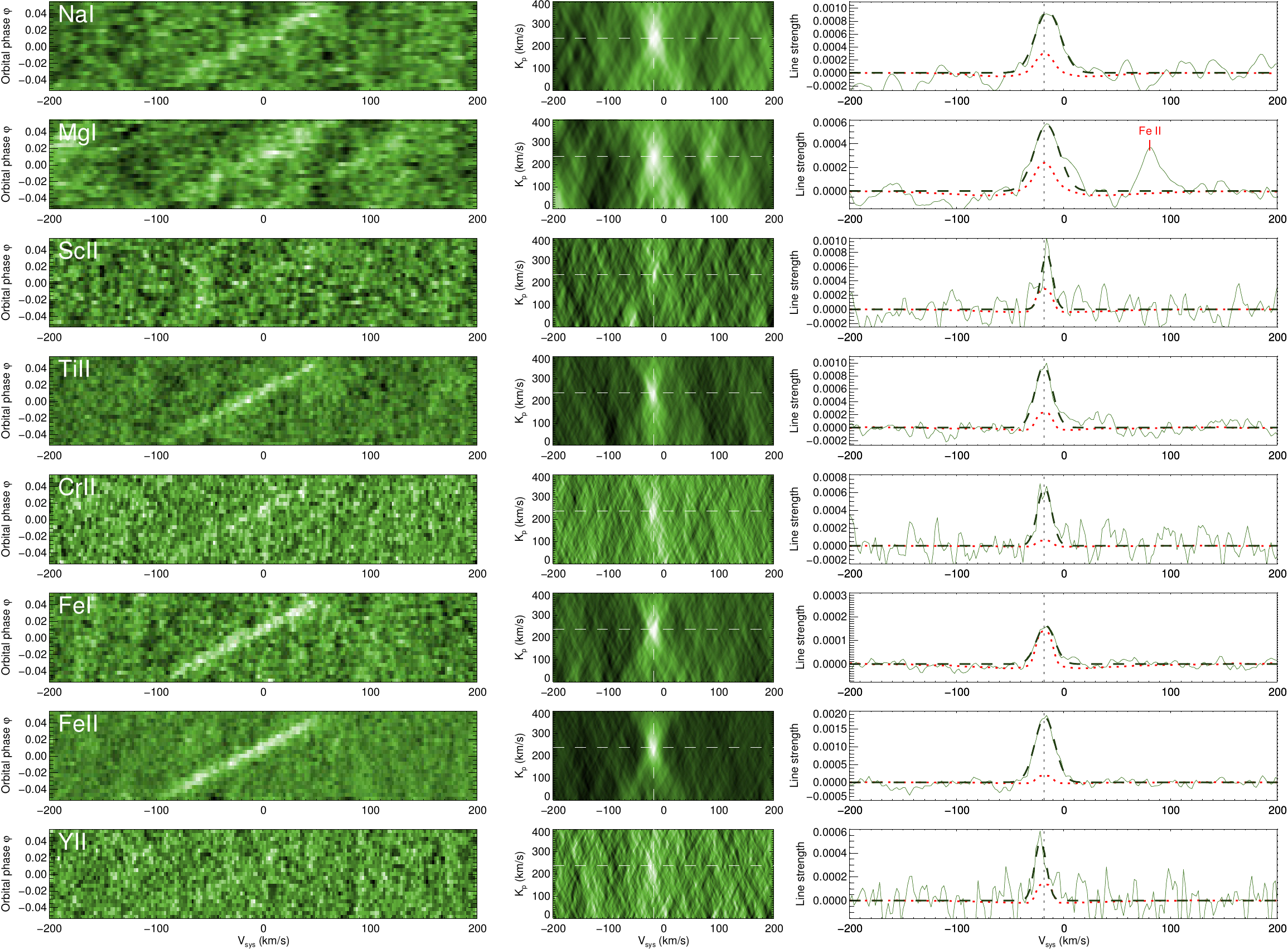}
   \caption{Cross-correlation signatures of \ion{Na}{i}, \ion{Mg}{i}, \ion{Sc}{ii}, \ion{Ti}{ii}, \ion{Cr}{ii}, \ion{Fe}{i}, \ion{Fe}{ii} and \ion{Y}{ii} averaged over both nights of observation. Each row of panels is associated with one species. The left panel in each row shows the the two-dimensional cross-correlation function $c(v,t)$, in which the signature of the transiting planet atmosphere has a slanted shape, as is most evident in the case of \ion{Fe}{i}, \ion{Fe}{ii} and \ion{Ti}{ii} that were previously reported in \citet{Hoeijmakers2018a}. The middle panel in each row shows the co-addition obtained by summing the in-transit exposures to the rest-frame velocity of the planet, assuming a range of orbital velocities, $K_p$. The white dashed lines indicate the expected systemic velocity, $v_{\textrm{sys}}$ (see section \ref{sec:velocity}) and orbital velocity (see section \ref{sec:stellarmass}). The right-most panel shows the one-dimensional cross-correlation function co-added in the rest-frame corresponding to the expected orbital velocity of the planet. The green line denotes the average cross-correlation over both observing nights. The dashed black line is a Gaussian fit to the profiles, with the resulting best-fit parameters provided in Table \ref{tab:fits}. The line-strength is converted to number of scale heights on the right-hand axis, computed at the level of the continuum ($T = 4,000 \textrm{ K}, \mu = 1.26 \textrm{ amu}, R_p = 1.891 R_J, M_p = 2.88 M_J, R_s = 2.362 R_{\odot}$).  The dotted red line is the signature of the injected model template. Apart from \ion{Fe}{i}, all observed absorption lines appear to be significantly deeper than predicted for a 4,000 K solar metallicity model in chemical equilibrium. The cross-correlation enhancement $\sim 80$ \kms redward of the \ion{Mg}{i} line is caused by a strong \ion{Fe}{ii} line, which is also indicated in Fig. \ref{fig:lines}.}%
   \label{fig:detections}
   \end{figure*}

Figure \ref{fig:all_correlations} shows the cross-correlation functions obtained when cross-correlating the observations of both nights with templates of all the selected species. We detect significant ($>5 \sigma$) excess absorption at the rest-frame velocity of the planet in the cross-correlation functions of \ion{Na}{i}, \ion{Mg}{i}, \ion{Sc}{ii}, \ion{Ti}{ii}, \ion{Cr}{ii}, \ion{Fe}{i}, \ion{Fe}{ii} and \ion{Y}{ii} (see Fig. \ref{fig:detections}). Of these species, \ion{Mg}{i}, \ion{Ti}{ii}, \ion{Fe}{i} and \ion{Fe}{ii} have previously been reported in the literature \citep{Hoeijmakers2018a,Cauley2018} and are confirmed by this analysis. In addition, we find tentative evidence of excess absorption by \ion{Ca}{i}, \ion{Cr}{i}, \ion{Co}{i}, and \ion{Sr}{ii} (see Fig. \ref{fig:all_correlations}), which appear to be located at the expected radial velocity of the planet and significant at the $>3\sigma$ level, but are irregularly shaped. Future observations will be needed to confirm or rule out the presence of these and possibly other absorbers in the spectrum of KELT-9 b. The cross-correlation templates used to detect \ion{Na}{i}, \ion{Mg}{i}, \ion{Sc}{ii}, \ion{Ti}{ii}, \ion{Cr}{ii}, \ion{Fe}{i}, \ion{Fe}{ii} and \ion{Y}{ii} are made available through the CDS \citep{Genova2000}. Templates of other species may be provided upon request.

   We fit the detected absorption lines by approximating them as Gaussian profiles. The error bars on the individual correlation values are empirically set to the standard deviation of $c(v)$ measured at velocities $200 < |v| < 1000$ \kms, away from the extrema of the radial velocity of the planet. The best-fit parameters obtained in this way are listed in Table \ref{tab:fits}. We perform a bootstrap analysis on the weakest detections of \ion{Y}{ii} and \ion{Sc}{ii} to determine the false-positive rate of these signatures. To this end, the CCF values away from the planet signal are permuted randomly 500,000 times before fitting a Gaussian, requiring a centroid position within 20 \kms of the expected radial velocity of the planet and a FWHM greater than 8 \kms. By fitting the tail of the resulting distribution as a power law, we find that signals with a strength corresponding to the detected \ion{Y}{ii} and \ion{Sc}{ii} lines are expected to occur randomly at rates of one in $1.6 \times 10^4$ and one in $1.7 \times 10^4$, respectively. To make sure that the detected signatures are not caused by weak residuals of the subtracted of the Doppler shadow, e.g. due to differences in the center-to-limb variation between neutrals and ions \citep{Yan2017}, we repeated the analysis by masking the radial velocities that are affected by the main shadow feature altogether. This did not significantly affect the strengths of the detected absorption lines.

 \begin{table*}
\caption{Best-fit parameters of the detected cross-correlation signatures. The profiles are fit by a Gaussian model using a Levenberg-Marquardt-algorithm. The error bar on each cross-correlation coefficient is approximated empirically by measuring the standard deviation at systemic velocities away from the orbital velocity of the planet ($200 < |v| < 1000$ \kms), from which the uncertainty intervals on the best-fit parameters are derived. This empirical treatment of the standard deviation is in line with previous studies that employ the cross-correlation technique, \citep[e.g][]{Hoeijmakers2015,Hoeijmakers2018b}. The second to fourth columns show the best-fit Gaussian amplitude, systemic velocity (equivalent to the planet rest-frame velocity at zero phase) and Full Width at Half Maximum (FWHM). These parameters are broadly consistent with those reported by \citet{Cauley2018}, who report line depths between 0.1\% and 0.5\% and FWHM widths between 20 and 40 \kms for \ion{Mg}{i}, \ion{Ti}{ii} and \ion{Fe}{ii}. The fifth column denotes the ratio between the best-fit amplitude and the line-strength of the injected model. The sixth column provides the altitude difference between the continuum and the line center in units of planetary radii. The last column expresses this altitude difference as the pressure at the line peak, assuming our isothermal hydrostatic atmospheric model which predicts that the continuum is formed by absorption by H$^-$ at a pressure of \continuumpressure. The last two columns also use the planetary and stellar parameters from \citet{Gaudi2017}. The uncertainty intervals on all parameters correspond to Gaussian $1\sigma$ intervals. These include the uncertainty on the fitting parameters but not the known uncertainties on the planetary and stellar parameters, nor systematic errors resulting from our choice of model.}
\label{tab:fits}      
\centering                          
\begin{tabular}{l|l l l l l l}        
\hline\hline                 
Species        & Amplitude  ($\times 10^{-3}$) & $v_{\textrm{sys}}$ (km/s) & FWHM (km/s) & Model discrepancy & $\Delta R_p$ & $ \log P (\textrm{bar})$ \\    
\hline                        

\ion{Na}{I}  & $0.952 \pm 0.065$ & $-14.8  \pm 1.0$  & $27.8 \pm 3.7$ & $3.02 \pm 0.20$ & $0.0581  \pm 0.0038$  & $-4.36  \pm 0.12$ \\
\ion{Mg}{I}  & $0.56  \pm 0.046$ & $-15.8  \pm 1.2$  & $27.5 \pm 4.3$ & $2.30 \pm 0.19$ & $0.0346  \pm 0.0028$  & $-3.61  \pm 0.10$ \\
\ion{Sc}{II} & $0.78  \pm 0.11$  & $-15.71 \pm 0.88$ & $12.7 \pm 2.1$ & $2.43 \pm 0.34$ & $0.0479  \pm 0.0066$  & $-4.04  \pm 0.21$ \\
\ion{Ti}{II} & $0.948 \pm 0.038$ & $-18.54 \pm 0.40$ & $19.3 \pm 1.1$ & $3.63 \pm 0.15$ & $0.0579  \pm 0.0023$  & $-4.436 \pm 0.071$ \\
\ion{Cr}{II} & $0.657 \pm 0.065$ & $-17.91 \pm 0.78$ & $15.7 \pm 2.0$ & $9.98 \pm 0.99$ & $0.0404  \pm 0.0040$  & $-3.80  \pm 0.13$ \\
\ion{Fe}{I}  & $0.163 \pm 0.006$ & $-16.9  \pm 0.35$ & $20.1 \pm 1.0$ & $1.05 \pm 0.04$ & $0.01019 \pm 0.00036$ & $-2.784 \pm 0.013$ \\
\ion{Fe}{II} & $1.831 \pm 0.042$ & $-17.56 \pm 0.25$ & $21.8 \pm 0.7$ & $8.67 \pm 0.20$ & $0.1091  \pm 0.0024$  & $-5.871 \pm 0.067$ \\
\ion{Y}{II}  & $0.50  \pm 0.08$  & $-21.6  \pm 1.1$  & $12.9 \pm 2.5$ & $3.23 \pm 0.54$ & $0.0309  \pm 0.0051$  & $-3.49  \pm 0.17$ \\

\hline                                   
\end{tabular}
\end{table*}

The fits summarized in Table \ref{tab:fits} signify a parametrization of a weighted average of the absorption lines of the species in this waveband over the duration of the transit. Injection of the templates into the data (as described in Section \ref{sec:injection}) allows these fitted profiles to be compared with the forward models of the atmosphere\footnote{Ideally, the spectra produced by the forward-models would be fit directly to the high-resolution spectra. However an efficient sampling of parameter space is currently not computationally tractable due to the size of the high-resolution spectra.}.

We caution that the measurement of a weighted average line function depends strongly on the weights attributed to each line. This point extends not only to the question of which lines are present in the cross-correlation template and their modelled relative depths, but also to the weights by which spectral bins and exposures are co-added, as well as any weighting or masking of individual spectral pixels. Comparison with an injected model provides consistency within the analysis, but these dependencies on the choices of weights make quantitative comparisons across the literature difficult. It could therefore be useful if standard templates defined over certain wavelength ranges be publicly available to the community. Regardless, our best-fit parameters are broadly consistent with the values found by \citet{Cauley2018} for their selections of lines of \ion{Mg}{i}, \ion{Ti}{ii} and \ion{Fe}{ii}.

\subsection{Line depth and atmospheric pressure}
\begin{figure}
\centering
\includegraphics[width=9cm]{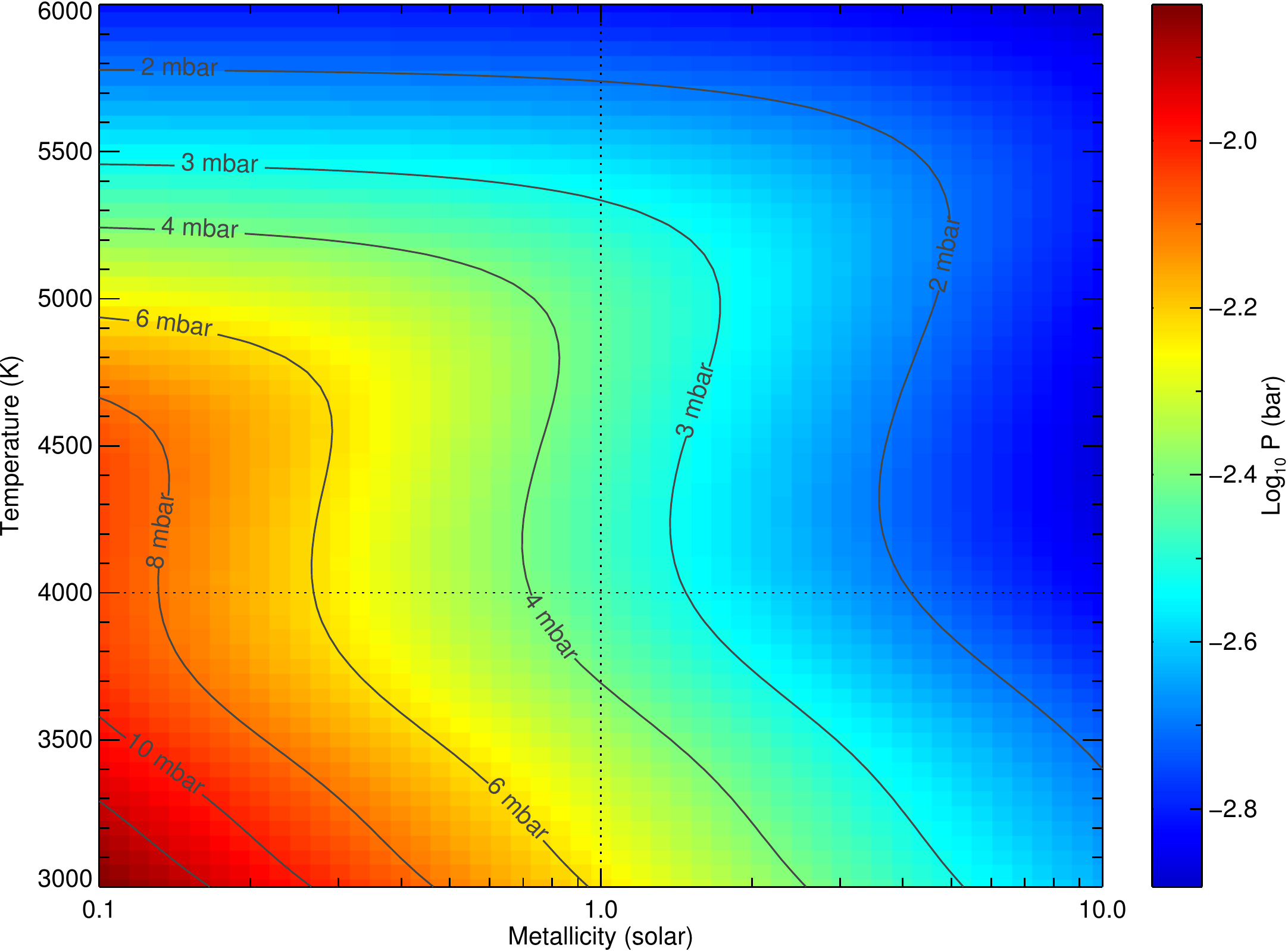}
\caption{The predicted pressure at which the atmosphere becomes optically thick to continuum radiation along the transit chord due to absorption by H$^-$, averaged over the HARPS waveband, as a function of atmospheric temperature and metallicity. The dashed lines indicate the metallicity and temperature chosen in this study, for which the continuum is predicted to become optically thick at a pressure of \continuumpressure. For solar metallicity, the continuum is formed between 3 mbar and 4 mbar for temperatures between 3,500 K and 5,500 K. For most of this parameter space, the continuum is formed between pressures of 2 mbar and 6 mbar. This indicates that the peak pressures obtained in Table \ref{tab:fits} are robust against errors in the assumed temperature and metallicity to within a factor of order unity.}%
\label{fig:contpressure}
\end{figure}

The best-fit parameters provided in Table \ref{tab:fits} reveal a discrepancy between the observed and the modeled line-depths. It appears that the presence of these strong ion lines is anomalous from the point of view of an isothermal equilibrium-chemistry model. Notably the \ion{Fe}{ii} profile spans $0.1091 \pm 0.0024$ planetary radii above the continuum under the assumption of a hydrostatic atmosphere, which is a factor of about 8.7 stronger than predicted by the model. Our model predicts that absorption by H$^-$ causes the atmosphere to become optically thick to continuum radiation at a pressure of \continuumpressure. This continuum pressure level can be used to break the normalization degeneracy of the transit radius that is inherent to transmission spectra of exoplanets \citep{Benneke2012,Griffith2014,Heng2017,Fisher2018}. We can therefore determine the pressure level corresponding to the peaks of the measured average absorption lines. These are reported in the last column of Table \ref{tab:fits}. In this way, we determine that the core of the \ion{Fe}{ii} line becomes optically thick at a pressure of $\log \frac{P}{1 \textrm{bar}} = -5.871 \pm 0.067$. As this constitutes a weighted average of many lines, the deepest \ion{Fe}{ii} lines necessarily become optically thick at even higher altitudes. We explore the model-dependency of the peak pressure levels by varying the temperature and metallicity, and find that the pressure level of the continuum vary from 2 mbar to 6 mbar for most of the parameter space between 3,000 to 6,000 K and 0.1 to 10$\times$ solar metallicity (see Fig. \ref{fig:contpressure}). We therefore conclude that the peak pressures obtained in Table \ref{tab:fits} are robust against errors in the assumed temperature and metallicity to within a factor of order unity.

Our chemistry model does not include photochemistry, so one could be tempted to explain the strong discrepancy between the observed line strengths and the injected model by the neglect of photo-ionization\footnote{In this high-temperature regime the term photo-chemistry loosely equates to photo-ionization because molecules are predominantly dissociated.}. However, the chemistry model suggests that thermal reactions by themselves are sufficient to nearly completely ionize \ion{Fe}{I} below pressures of \continuumpressure (see Fig. \ref{fig:abundances}). An additional source of ionization could therefore act to further deplete the neutral \ion{Fe}{i} population, but this would only marginally increase that of \ion{Fe}{ii}. Therefore we conclude that deviations from the assumed ion chemistry cannot easily explain the observed discrepancy in the absorption lines of ions.

A similar reasoning applies to vertical mixing, which is neglected in our model because of the short thermal timescale at a temperature of 4,000 K. The abundances would be quenched if the vertical mixing timescale would be short at the pressures probed in the transmission spectrum, and this would result in abundance profiles that are constant with pressure/altitude \citep[see e.g.][]{Tsai2018}. Although this effect would work to increase the abundances of neutral elements at higher altitudes, the abundances of singly ionized species are already nearly constant between pressures of 0.1 to $10^{-14}$ bars (see Fig. \ref{fig:abundances}). Quenching can therefore not be invoked to significantly increase the line depths of the ionized species, and the same reasoning applies to horizontal transport. However, as disequilibrium chemistry driven by transport may be relevant for the abundances of the neutral trace species \citep{Mendonca2018}, future studies should treat the effects of three-dimensional transport.

A fourth simplifying assumption in our model is that the abundance of metals in the atmosphere corresponds to solar metallicity, equivalent to the metallicity of the star \citep{Gaudi2017}. If the metallicity of the planet is significantly enhanced, this would increase the optical depth inside the absorption lines of all elements. However, the altitude at which the atmosphere becomes optically thick in transmission scales with the logarithm of the elemental abundance \citep{Lecavelier2008,deWit2013,Heng2017}. Increasing the line-depth by a factor of eight to ten would require an increase in the metallicity of four to five orders of magnitude. For a high metallicity atmosphere, the scale height would be decreased due to the higher mean particle weight, reducing the line depths rather than increasing them. In addition, gas giant exoplanets with masses greater than Jupiter are not expected to have metallicities that are orders of magnitude greater than stellar \citep[e.g.][]{Arcangeli2018}.

Fifthly, the model assumes an isothermal atmosphere. If parts of the atmosphere are significantly hotter than 4,000 K, this would increase the scale height and therefore inflate the observed absorption lines. In previous studies, this effect has been used to measure the temperature-pressure profile using the line profile of the strong sodium D-doublet \citep[see e.g.][]{VidalMajar2011,Heng2015,Wyttenbach2015,Pino2018a}. Although the isothermal assumption is not likely to accurately describe the atmosphere over a wide range of pressures, explaining the observed discrepancy of the observed and predicted \ion{Cr}{ii} and \ion{Fe}{ii} lines by invoking higher temperatures would imply an unrealistically strong thermal inversion: To increase the scale-height $H$ by a factor of ten, the temperature of the isothermal atmosphere would need to be increased to $\sim$40,000 K because line-depth in transmission scales approximately linearly with $H$.

A sixth assumption is that of local thermodynamic equilibrium, resulting in the atomic level populations being determined by thermal collisions. At the low pressures where some of the stronger lines are formed, the atomic energy levels may be significantly populated via non-thermal processes. It is not immediately evident how deviations from LTE distributions of the level populations would affect the observed spectra of the different species, and this is a topic deserving of further theoretical study that is beyond the scope of this paper \citep[along the lines of e.g.][]{Huang2017,Oklopcic2018}.

Deep absorption lines by neutral and ionized elements have previously been observed in the transmission spectra of other hot Jupiters, where they are assocated with outflowing material that is escaping from the planet atmosphere \citep[][Allart 2019, in prep]{VidalMajar2003,VidalMajar2004,BenJaffel2010,Fossati2010,Linsky2010,Schlawin2010,VidalMajar2011,Ehrenreich2012,Haswell2012,Kulow2014,Bourrier2015,Ehrenreich2015,Spake2018,Allart2018,Nortmann2018,Salz2018}.
An evaporation outflow effectively enhances the density of the atmosphere at higher altitudes, lifting absorbing species above what would be expected from a hydrostatic atmosphere. KELT-9 b is also believed to possess an outflowing envelope, as evidenced by strong H-$\alpha$ absorption in the optical transmission spectrum \citep{Yan2018}. Therefore, we interpret these deep ion lines as originating from the evaporation outflow that increases their density at higher altitudes than expected from our model that is in hydrostatic equilibrium. The strength of such a potential outflow is still a topic of discussion \citep[see e.g.][]{Fossati2018}, and future studies will be directed at modeling this outflow over the pressure ranges probed in the high-resolution transmission spectrum.

\subsection{Individual lines of \ion{Fe}{ii}, \ion{Mg}{i} and \ion{Na}{i}}
Strong absorption lines of \ion{Mg}{i} and \ion{Fe}{ii} have already been observed directly in the transmission spectrum of KELT-9 b by \citet{Cauley2018} using the PEPSI spectrograph on the 8.2-m Subaru telescope. We therefore used these HARPS-N observations to obtain the transmission spectrum of KELT-9 b using the same methods as described by \citet{Wyttenbach2015}, but using a low-order polynomial to normalize the continuum \citep[similar to][]{Allart2017}, and correcting for the Doppler-shadow in the same way as was done for the cross-correlation functions (see Section \ref{sec:obs}).

\begin{figure*}
   \centering
   \includegraphics[width=18cm]{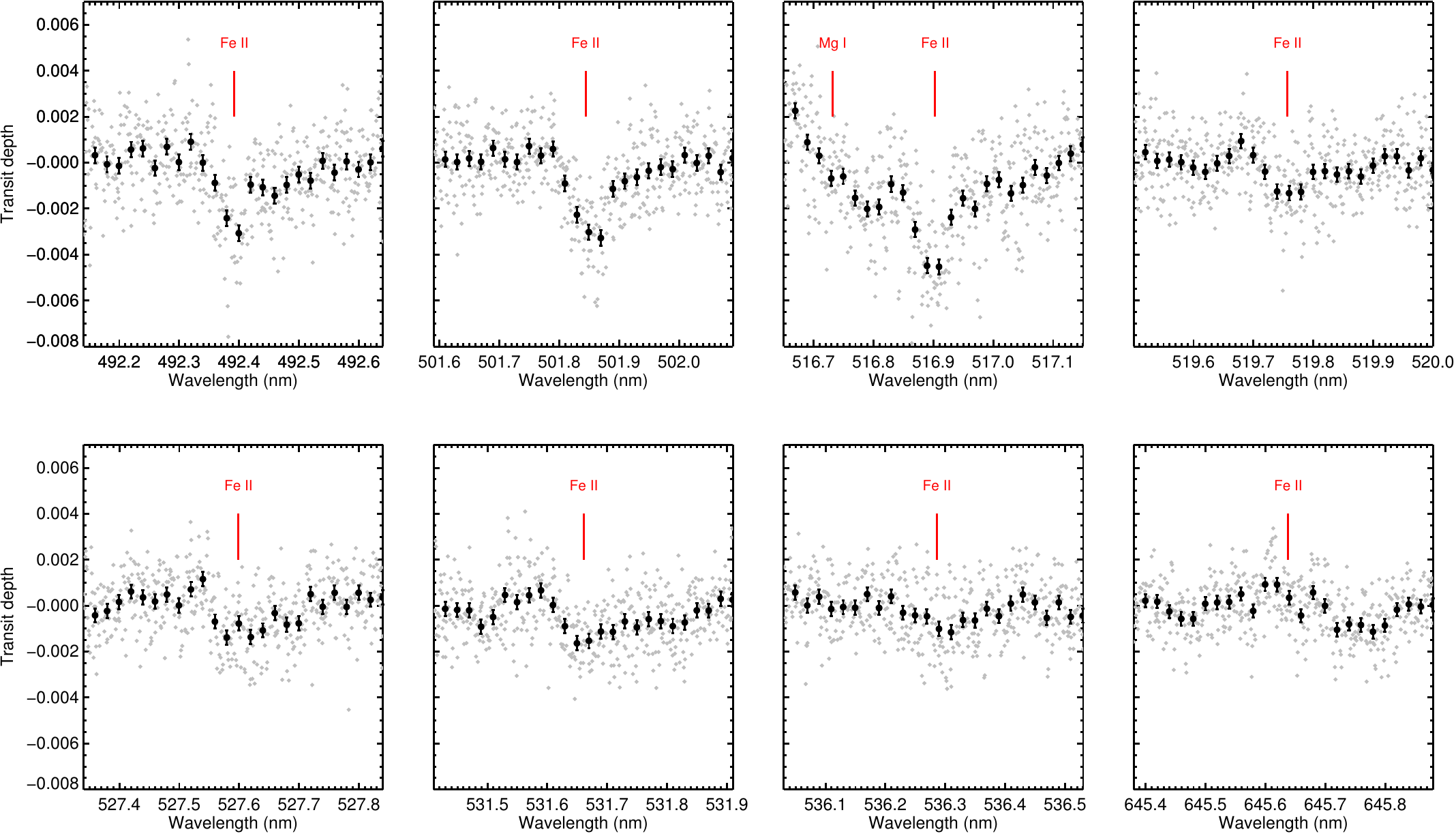}
   \caption{The transmission spectrum of KELT-9 b at the location of eight strong lines of \ion{Fe}{ii} that were observed by \citet{Cauley2018}. The gray points show the transmission spectrum at the native sampling of the HARPS-N instrument, while the black points are binned by 20 points. The transmission spectrum is obtained following the procedure of \citet{Wyttenbach2015}, and combining all in-transit exposures of the two nights. The Doppler-shadow is removed by fitting the model described in Section \ref{sec:obs} at the expected locations of the individual lines. The presence of these lines appears to be confirmed by our HARPS-N observations, but non-Gaussian systematic noise occurs at the $\sim 10^{-3}$ level. This hinders the identification of all but the strongest \ion{Fe}{ii} lines.}%
   \label{fig:lines}
   \end{figure*}

\begin{figure*}
   \centering
   \includegraphics[width=18cm]{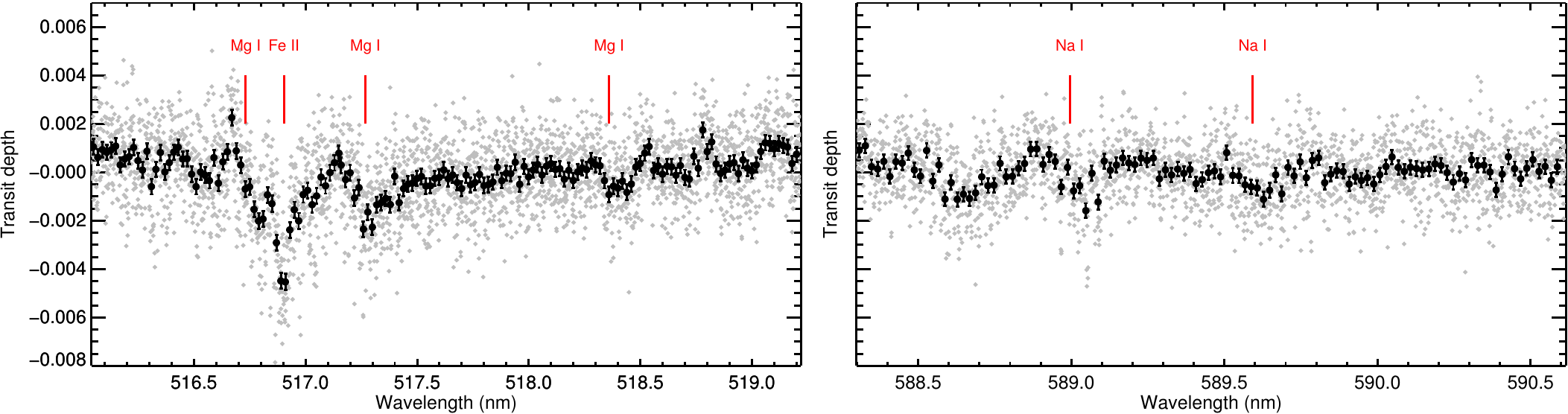}
   \caption{Similar to Fig. \ref{fig:lines}, this figure shows the transmission spectrum of KELT-9 b at the location of the \ion{Mg}{i} triplet that was observed by \citet{Cauley2018}, and the \ion{Na}{i}-D doublet, which was not covered by the observations by \citet{Cauley2018}. The systematic noise at the $\sim 10^{-3}$ level hampers the detection of these individual lines.}%
   \label{fig:MgNalines}
   \end{figure*}

Figures \ref{fig:lines} and \ref{fig:MgNalines} show the transmission-spectrum obtained in this way, at the location of eight strong \ion{Fe}{ii} lines also observed by \citet{Cauley2018}, the \ion{Mg}{i} triplet and the \ion{Na}{i}-D doublet\footnote{Single lines and simple line systems like doublets and triplets are more amenable for a targeted analysis that is applied directly to the transmission-spectrum \citep[like e.g.][]{Wyttenbach2015} rather than cross-correlation, because cross-correlation relies on the co-addition of many individual lines.}. The strongest lines of \ion{Fe}{ii} are detected individually in the transmission spectrum, but systematic variations are present on the $10^{-3}$ level\footnote{The origin of these ystematic variations is presently unknown and may be related to the malfunction of the ADC. The magnitude of the variations exceeds the magnitude of the Doppler shadow, meaning that inaccuracies in the removal of the Doppler shadow effect cannot (fully) account for this systematic noise.}. Excess absorption also appears around the \ion{Na}{i} doublet, but the detection of these lines in the transmission spectrum is not robust. This highlights the advantage of using cross-correlation, in which many lines (if available) are co-added to diminish the photon noise and other noise sources that are not correlated with the target spectrum. In the optical, \ion{Na}{i} has tradionally only been targeted through its strong resonant doublet, but at high temperatures additional transitions are also common, producing the multitude of lines that were combined in the present analysis, yielding a detection that is significantly stronger than would have been possible using only the doublet.

In addition, neutral aluminium (\ion{Al}{i}) and ionized calcium (\ion{Ca}{ii}) have strong doublets near 390 nm (see Fig. \ref{fig:spectra}). However these were not detected because the noise at the blue edge of the HARPS-N waveband is naturally high, aggravated by the malfunction of the ADC that caused a strong loss of flux at short wavelengths in the second night of observations. Besides these two doublets, the spectra of most other species appear to be more rich at blue optical and UV wavelengths (see Fig. \ref{fig:spectra}) and we propose that future spectroscopic observations at blue/NUV wavelengths (notably with HST/STIS) have a strong potential to reveal additional absorbers in the transmission spectrum of the atmosphere of KELT-9 b.

\subsection{The systemic velocity and high-altitude winds}\label{sec:velocity}
The HARPS-N spectrograph is routinely used to monitor the radial velocity of stars with a long-term precision at the m s$^{-1}$ level \citep{Cosentino2012}. For this purpose the pipeline automatically cross-correlates the spectrum of the target star with a standard binary mask, in this case a mask tailored to the spectrum of an A0 main sequence star (Nielsen et al. in prep). Using the cross-correlation functions provided by the pipeline, we measure the radial velocity of the star in the exposures obtained before and after the transit\footnote{The in-transit exposures are left out as to prevent the presence of the planet to bias the velocity measurement}. For the two nights, this yields measurements of the systemic radial velocity of $-17.63\pm0.14$ \kms and $-17.88\pm0.16$ \kms respectively.

\citet{Gaudi2017} report a systemic velocity of $-20.565 \pm 0.1$ \kms, which is significantly discrepant from our measurements. The reason for such a discrepancy at the \kms level is not immediately evident, but may well be explained by differences in the cross-correlation masks used to derive the radial velocity, as well as the radial-velocity zero-point of the spectrographs. For the purpose of this study we adopted a systemic velocity of $-17.74 \pm 0.11$ \kms as the average of our two measurements.

The strong signatures of \ion{Na}{i}, \ion{Mg}{i}, \ion{Ti}{ii}, \ion{Fe}{i} and \ion{Fe}{ii} in this survey appear to have symmetrical line profiles at velocities that are consistent with the measured systemic velocity to within $3 \sigma$. Especially the \ion{Fe}{ii} line, although spanning a large range of altitudes, is consistent with the systemic velocity as it deviates from the planet rest frame velocity by \maxfeiiwindspeed \kms. Similarly, the \ion{Fe}{i} line is shifted by \maxfeiwindspeed \kms. Our detections therefore do not indicate the presence of a strong net day-to-night-side wind (the second circulation regime of \citet{Showman2013}), nor a tail-like flow of the envelope material over the range of pressures probed by these lines - i.e. from the millibar to the microbar level.

\subsection{The stellar mass}\label{sec:stellarmass}
\begin{figure}
  \centering
  \includegraphics[width=9cm]{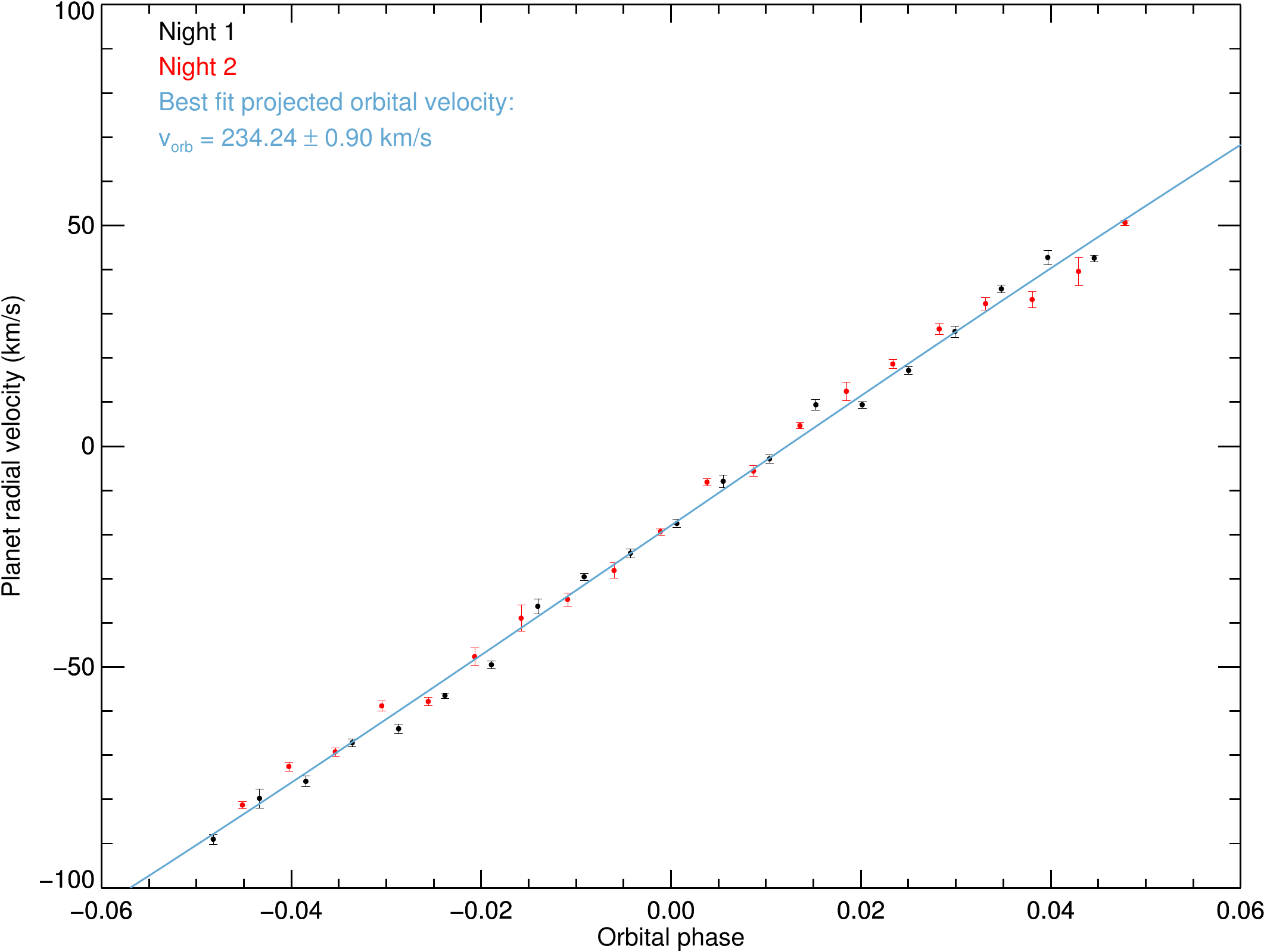}
  \caption{The radial velocity of the planet as a function of orbital phase. It is obtained by fitting the average \ion{Fe}{ii} line as a Gaussian in each exposure of each of the two nights. The blue line signifies our fit of Eq. \ref{eq:vorb}, which yields a best-fit projected orbital velocity of \orbitalvelocity \kms.}%
   \label{fig:fitrv}
\end{figure}

The detection of the absorption lines in the transmission spectrum of the planet yields a measurement of the radial velocity of the planet as a function of time during transit. Assuming a circular orbit and the known orbital period, this yields the projected orbital velocity of the planet. This was first done by \citet{Brogi2012} who measured the true orbital velocity of the non-transiting hot Jupiter $\tau$ Boo b, which was first known only up until a factor $\sin i$ because the inclination of the orbit was unknown. Assuming a circular orbit, we fit the radial velocity $v(t)$ of the peak of the cross-correlation function of \ion{Fe}{ii} in each exposure during the transit to obtain the orbital velocity $v_{\textrm{orb}}$, up to the projection angle $\sin i$:

\begin{equation}\label{eq:vorb}
  v(t) = v_{\textrm{orb}} \sin \left( 2 \pi \frac{t - t_0}{P} \right) \sin i \ ,
  \end{equation}

where $t - t_0$ and $P$ are the time from transit center time and the orbital period as provided by \citet{Gaudi2017}. The resulting fit is shown in Fig. \ref{fig:fitrv}, and yields a projected orbital velocity of $\overline{v_{\textrm{orb}}} = v_{orb} \sin i = 234.24 \pm 0.90$ \kms.

Because KELT-9 b is a transiting planet, knowledge of the orbital inclination, the orbital period and the orbital velocity may be used to retrieve the true mass of the star. Rewriting Keplers third law for a circular orbit, the stellar mass $M_*$ is given by

\begin{equation}\label{eq:stellarmass}
  M_* = \frac{P}{2\pi G} v_{\textrm{orb}}^3 - m_p = \frac{P}{2 \pi G} \left( \frac{\overline{v_{\textrm{orb}}}}{\sin i}\right)^3 - m_p  \ ,
\end{equation}

where $m_p$ is the mass of the planet. Filling in the measured orbital velocity and the inclination, orbital period and planet mass reported by \citet{Gaudi2017}, we obtain a revised measurement of the stellar mass of $M_* = \stellarmass$. The uncertainty on this value is approximately 10 times lower than reported by \citet{Gaudi2017}, and takes into account the uncertainties on $P$, $i$ and $m_p$. The stellar density and planet-to-star radius ratio were obtained by \citet{Gaudi2017} as direct observables from the shape and depth of the transit lightcurve \citep{Seager2003}, who reported values of $\rho_* = 0.2702 \pm 0.0029$ \gcm and $\frac{R_p}{R_*} = 0.08228 \pm 0.00043$. We are therefore able to revise the stellar radius to $R_* = \stellarradius$ and the planet radius to $R_p = \planetradius$. Finally, with the revised stellar mass we also revise the mass of the planet, which is derived from the radial velocity semi-amplitude $K$ of the star provided by \citet{Gaudi2017}, assuming a circular orbit:

\begin{equation}
  m_p = K \left(\frac{P}{2\pi G}\right)^\frac{1}{3} \frac{M_*^\frac{2}{3}}{\sin i}
\end{equation}

which yields a mass of $\planetmass$ \footnote{The planet mass $m_p$ recurs in Eq. \ref{eq:stellarmass} for the mass of the star, but the adjustment in the planet mass compared to \citet{Gaudi2017} is so small that it has no effect on the initial outcome for $M_*$. Had the planet been significantly heavier, $M_*$ and $m_p$ would have been derived iteratively.}.

\section{Conclusion}
This paper presents a survey of the transmission spectrum of KELT-9 b for absorption signatures of 75 atomic species and their ions, as observed by the high-resolution echelle spectrograph HARPS-N. Of these 75 species, 18 neutrals and 23 ions are expected to have significant line absorption in the HARPS-N waveband. We modeled the absorption spectra of these species using the \texttt{FastChem} code \citep{Stock2018} to model the expected composition of the atmosphere at 4,000 K, and used these model spectra as cross-correlation templates to co-add the individual absorption lines of these species into average line-profiles. By doing so, we are able to report on new detections of \ion{Na}{i}, \ion{Sc}{ii}, \ion{Cr}{ii} and \ion{Y}{ii} in the atmosphere of KELT-9 b, while confirming previous detections of \ion{Mg}{i}, \ion{Ti}{i}, \ion{Fe}{i} and \ion{Fe}{ii} \citep{Hoeijmakers2018a,Cauley2018}. In addition, we find cross-correlation enhancements of \ion{Ca}{i}, \ion{Cr}{i}, \ion{Co}{i}, and \ion{Sr}{ii} near the expected orbital velocity of the planet but these have low significance and/or non-Gaussian profiles. These tentative signatures will require further observations to verify.

The injection of the model templates into the data allows the strength of the cross-correlation functions to be compared to the models. From this we conclude that equilibrium chemistry at solar metallicity and 4,000 K comes close to adequately explaining the observed line-depths of \ion{Fe}{i} and \ion{Sc}{ii}. However the strong absorption lines of the other species appear to be anomalous, especially those of \ion{Cr}{ii} and \ion{Fe}{ii}, which are both underpredicted by a factor of eight to ten. Our atmospheric model predicts that the atmosphere becomes opaque to continuum radiation at a pressure of \continuumpressure. We could therefore derive the pressure level corresponding to the observed line-depths and find that these occur between the mbar and microbar levels. We tentatively ascribe the discrepancy between our isothermal, hydrostatic thermal equilibrium model and the data to the presence of a hydrodynamic outflow that lifts the absorbing species to higher altitudes. The presence of an escaping envelope surrounding KELT-9 b has already been confirmed by recent observations of the H-$\alpha$ line by \citet{Yan2018}.

From the cross-correlation performed by the HARPS-N pipeline, we constrain the systemic velocity of the KELT-9 system to $-17.74 \pm 0.11$ \kms, significantly deviant from the value of $-20.565 \pm 0.1$ \kms reported by \citet{Gaudi2017}. The observed absorption lines in the transmission spectrum of the planet atmosphere do not show a significant net blue-shift from the systemic velocity that would indicate a global day-to-night side wind or a tail-like structure of the envelope material at the pressures probed.

We measure the variation of the radial velocity of the planet during transit using the strong \ion{Fe}{ii} lines, and find a projected orbital velocity of \orbitalvelocity \kms. Treating the system like a spectroscopic binary and using the transit parameters reported by \citet{Gaudi2017}, we revise the stellar mass and radius to $M_* = \stellarmass$ and $R_* = \stellarradius$, and the planet mass and radius to $m_p= \planetmass$ and $R_p = \planetradius$.

\begin{acknowledgements}
This project has received funding from the European Research Council (ERC) under the European Union’s Horizon 2020 research and innovation programme (projects Four Aces and EXOKLEIN with grant agreement numbers 724427 and 771620, respectively). This work has been carried out in the framework of the PlanetS National Centre of Competence in Research (NCCR) supported by the Swiss National Science Foundation (SNSF). Based on observations made with the Italian Telescopio Nazionale Galileo (TNG) operated on the island of La Palma by the Fundación Galileo Galilei of the INAF (Istituto Nazionale di Astrofisica) at the Spanish Observatorio del Roque de los Muchachos of the Instituto de Astrofisica de Canarias. This project has received funding from the European Union's Horizon 2020 research and innovation programme under grant agreement No 730890 (OPTICON). This material reflects only the authors views and the Commission is not liable for any use that may be made of the information contained therein. A.W acknowledges the support of the SNSF by the grant number P2GEP2\_178191.
\end{acknowledgements}

\begin{appendix}

 \section{The weighted cross-correlation}\label{app:ccv}
 The cross-correlation $c(v,t)$ provides an independent measurement of the average line profile for each exposure in the time-series of the observations. Hence, $c(v,t)$ is a function of both the time of observation (or equivalently the orbital phase of the planet) and the radial velocity to which the template is shifted. This yields the two-dimensional cross-correlation maps that are regularly presented in studies that use the cross-correlation technique, e.g. Figures 5, 8 and 12 in \citet{Brogi2018}, \citet{Birkby2017} and \citet{Hoeijmakers2018b} respectively. $c(v,t)$ is computed independently for each of the wavelength bins. In our study these are 20 nm-wide sections of the stitched 1D spectra as produced by the HARPS-N pipeline, but they could also be the individual spectral orders of the echelle spectrograph. To combine these, we first take the weighted mean of $c(v)$ over the $N_b = 15$ bins to produce the two-dimensional cross-correlation function over the entire spectral range, followed by a second weighted mean over the $N_e = 19$ exposures in-transit, yielding a one-dimensional cross-correlation $C(v)$ function that corresponds to the mean line profile over the entire dataset:

\begin{equation}
C(v) =\sum_{k=0}^{N_e} \sum_{j=0}^{N_b} \sum_{i=0}^{N} \frac{x_{ijk}T_{ij}(v)w_{j}e_k}{\sum_{k=0}^{N_e}e_k \sum_{j=0}^{N_b}w_{j}  \sum_{i=0}^{N}T_{ij}(v)}
\end{equation}

where $e_k$ is the weight assigned to the value of $c(v)$ in exposure $k$, $w_j$ is the weight assigned to $c(v)$ of each of the 15 bins, $x_{ijk}$ is spectral data point $i$ in bin $j$ and exposure $k$, and $T_{ij}(v)$ is the template corresponding to each of these $x_{ijk}$ shifted to radial velocity $v$.

The weights $w_j$ and $e_k$ are based on the signal-to-noise of the signal retrieved when injecting the template $T$ into the data, broadened by the expected rotational velocity of the planet assuming tidal locking (see section \ref{sec:transmission_spectra}). The weights are thus effectively optimized to retrieve a transmission spectrum of the planet that corresponds to $T$, and this is done for each species independently. This assures that the spectral pixels are all co-added based on not only the expected position and strengths of the absorption lines (as follows from the forward-model of the atmosphere), but also implicitly accounting for the wavelength-dependence of noise over the spectral range. Regions with high flux and deep planetary absorption lines are thus naturally favored over regions where the noise is higher (such as at the edges of spectral orders, at blue wavelengths and in deep stellar absorption lines), or where the planet has fewer or shallower lines.

The cross-correlation of the injected template is denoted as $C_I$, and its absolute signal is obtained by subtracting $C_I - C$. The peak signal-to-noise of this cross-correlation signal occurs at the radial velocity of the planet at which the template $T$ was injected (corresponding with the true expected radial velocity of the planet in-transit), and is denoted as $s_{jk}$, for each bin $j$ and exposure $k$. $e_k$ is now obtained by taking the mean of $s^2$ over $j$, and $w_j$ is obtained by taking the mean of $s^2$ over $k$.

\section{Notes on nomenclature}
As already noted by \citet{Pino2018}, the functional form of what is referred to as 'cross-correlation' varies between fields, and the distinction is not always made clear in the exoplanet literature. \citet{Snellen2010,Brogi2012,Hoeijmakers2015} a.o. use the Pearson correlation coefficient, which is similar to Eq. \ref{eq:ccv}, but normalizes $x_i$ and $T_i$ by subtracting their mean, and carries an additional normalization over $x_i$ in the denominator. As such, this cross-correlation coefficient is unitless and bounded between -1 and 1. Its magnitde can therefore not be related directly to the amplitude of the target spectral lines, because this information has been normalized out. Relating the amplitude of the Pearson cross-correlation to a physical line-stregth therefore relies on injection (as in e.g. \citet{Hoeijmakers2015}) or subtraction (as in e.g. \citet{Brogi2017}) of the model template to compare the real and injected cross-correlation amplitudes.

\citet{Allart2017,Hoeijmakers2018a} as well as the present work refer to Eq. \ref{eq:ccv} as the cross-correlation, although it is functionally identical to a weighted mean of spectral pixels, and more closely related to the cross-covariance. This function preserves the physical unit of the data (in the present case: transit depth), allowing a direct interpretation of the amplitude of the cross-correlation signal as the average line-depth \citep{Pino2018}. The cross-covariance can also be incorporated into an expression of the likelihood function $L$, allowing it to be used as a goodness-of-fit parameter \citep{Brogi2018}.

\section{The full survey}
\begin{figure*}
  \caption{Modeled spectra of all neutral and singly ionized species in our survey, with the atomic number increasing downwards. The spectra are computed at solar metallicity and a temperature of 4,000 K, with a continuum set by H$^-$. The y-axis denotes transit depth, with unity equal to the out-of-transit flux of the star. The shaded gray areas demarcate the edges of the HARPS-N waveband. The alternating shaded colours indicate the 20 nm wide bands for which the cross-correlation analysis is performed independently. Green/lime highlighted species are species with significant expected line opacity and that were selected for cross-correlation. Species for which no spectrum is plotted were lacking in line-list data.}
  \smallskip
  \includegraphics[angle=90,width=16.5cm]{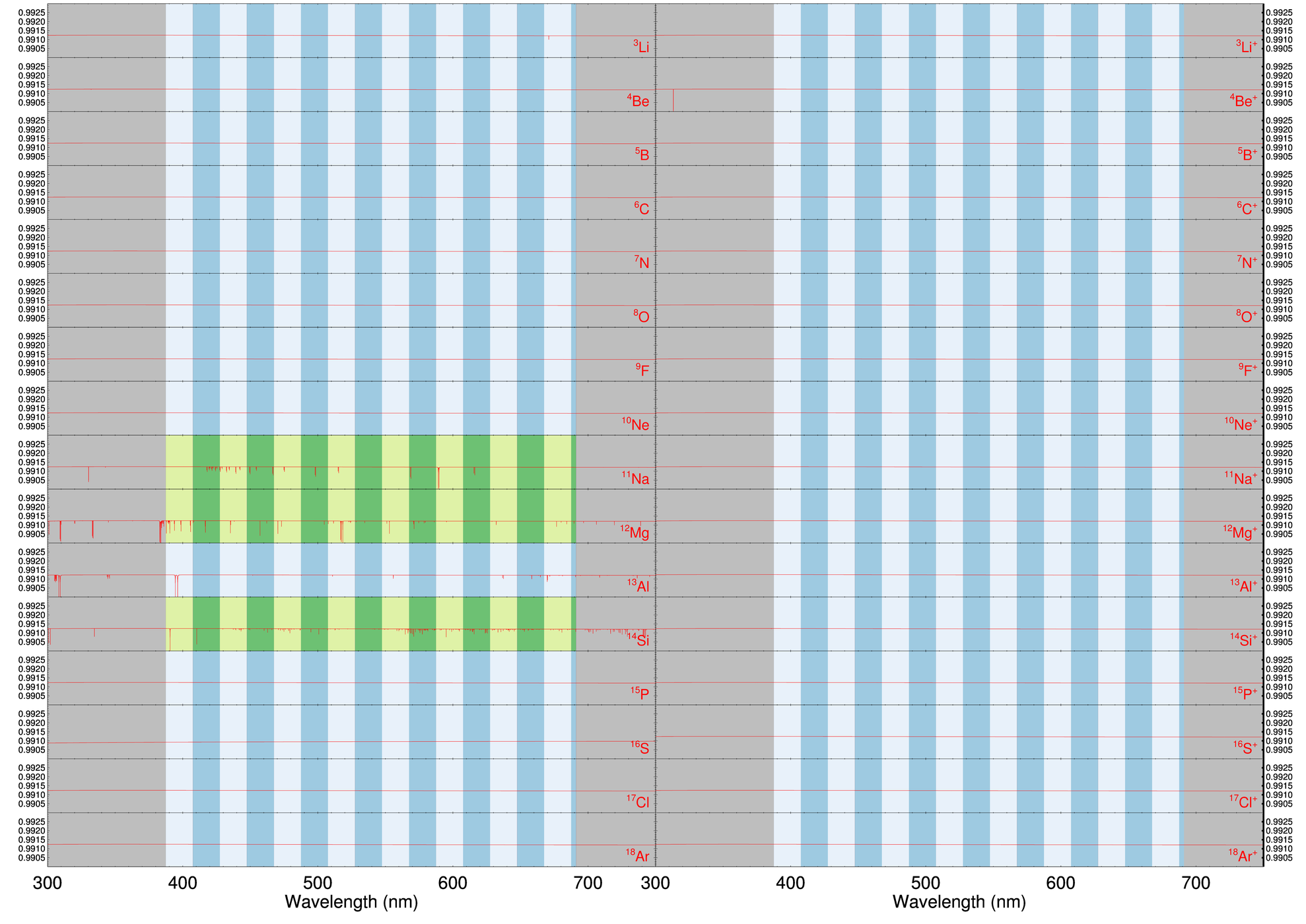}
  \label{fig:spectra}
\end{figure*}

Besides the detected species shown in Fig. \ref{fig:detections}, in Fig. \ref{fig:spectra} we present the modelled absorption spectra of all species between atomic number 3 and 78 for which line-list data was available. We chose not to include elements with atomic number above 78 because the modelled spectra were mostly devoid of lines and line-list data was sparse. Fig. \ref{fig:all_correlations} shows the cross-correlation signature of each of the species in the same way as Fig. \ref{fig:detections}. Table \ref{table:DSfitting} shows the scaling factors by which the model of the Doppler shadow was multiplied before subtraction. Finally, Table \ref{table:FastChemData} shows the coefficients of our fits to the mass action constants used to compute the equilibrium chemistry model of the atmosphere using \texttt{FastChem}.

\begin{figure*}
  \includegraphics[angle=90,width=18cm]{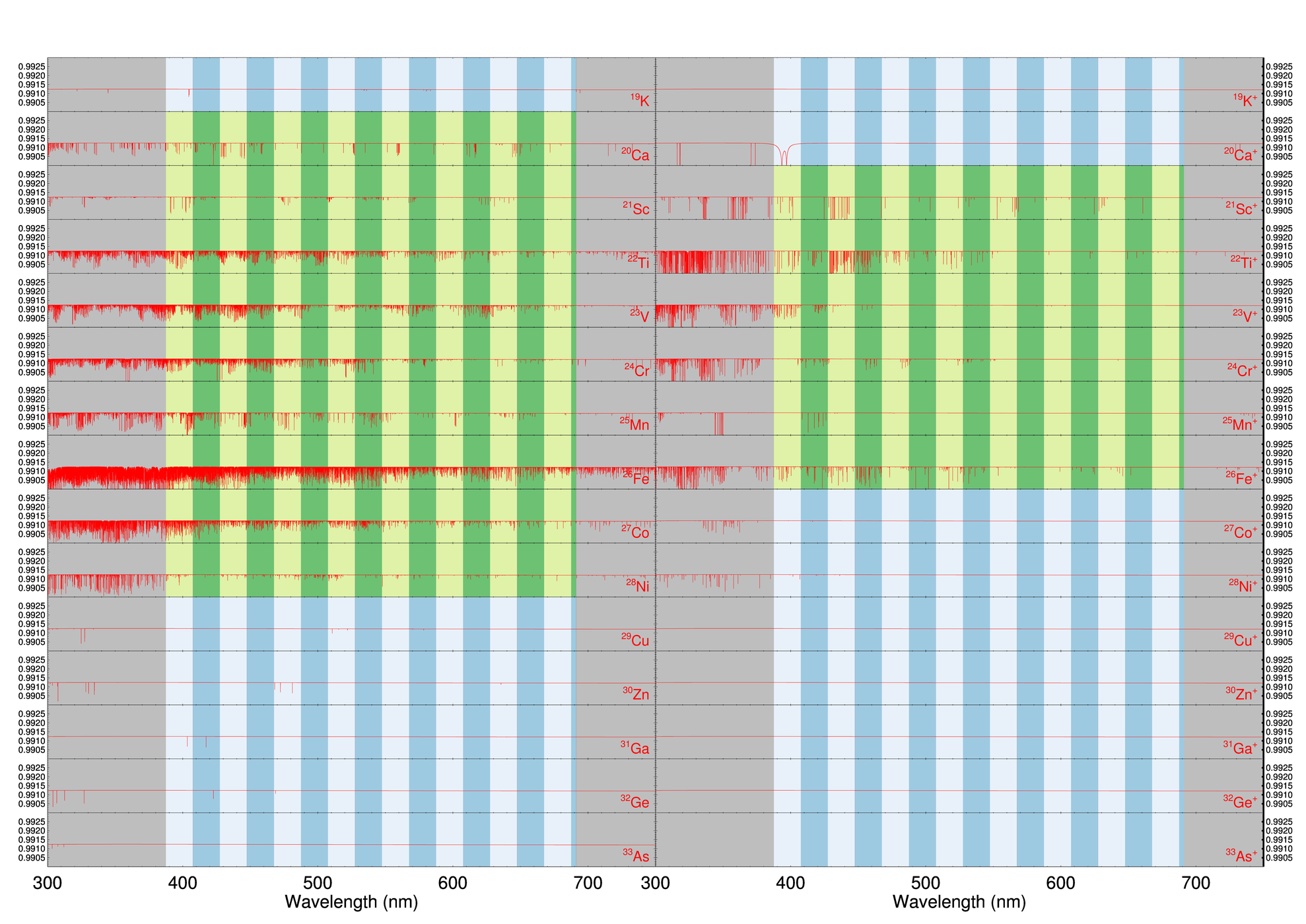}
\end{figure*}

\begin{figure*}
  \includegraphics[angle=90,width=18cm]{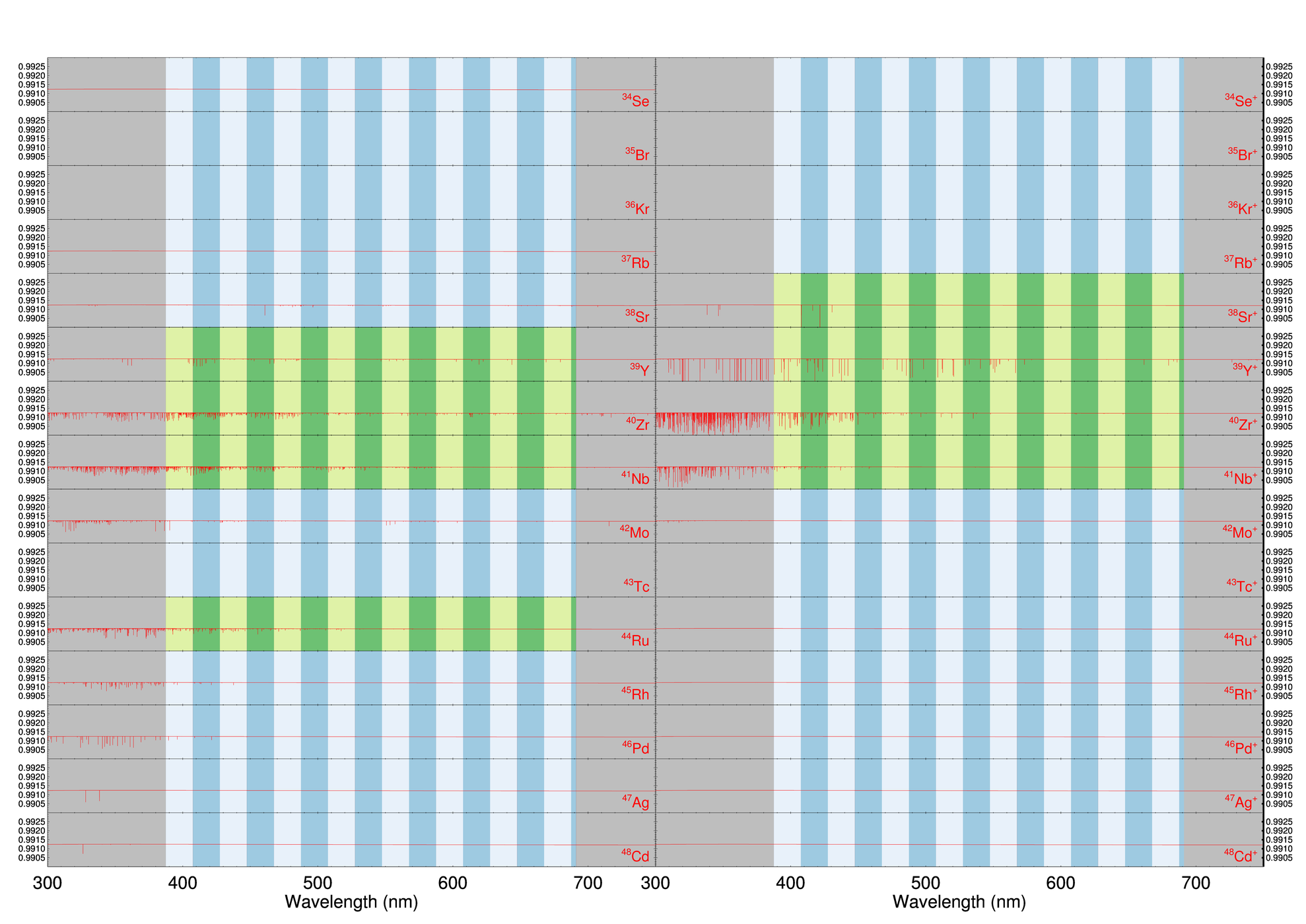}
\end{figure*}

\begin{figure*}
  \includegraphics[angle=90,width=18cm]{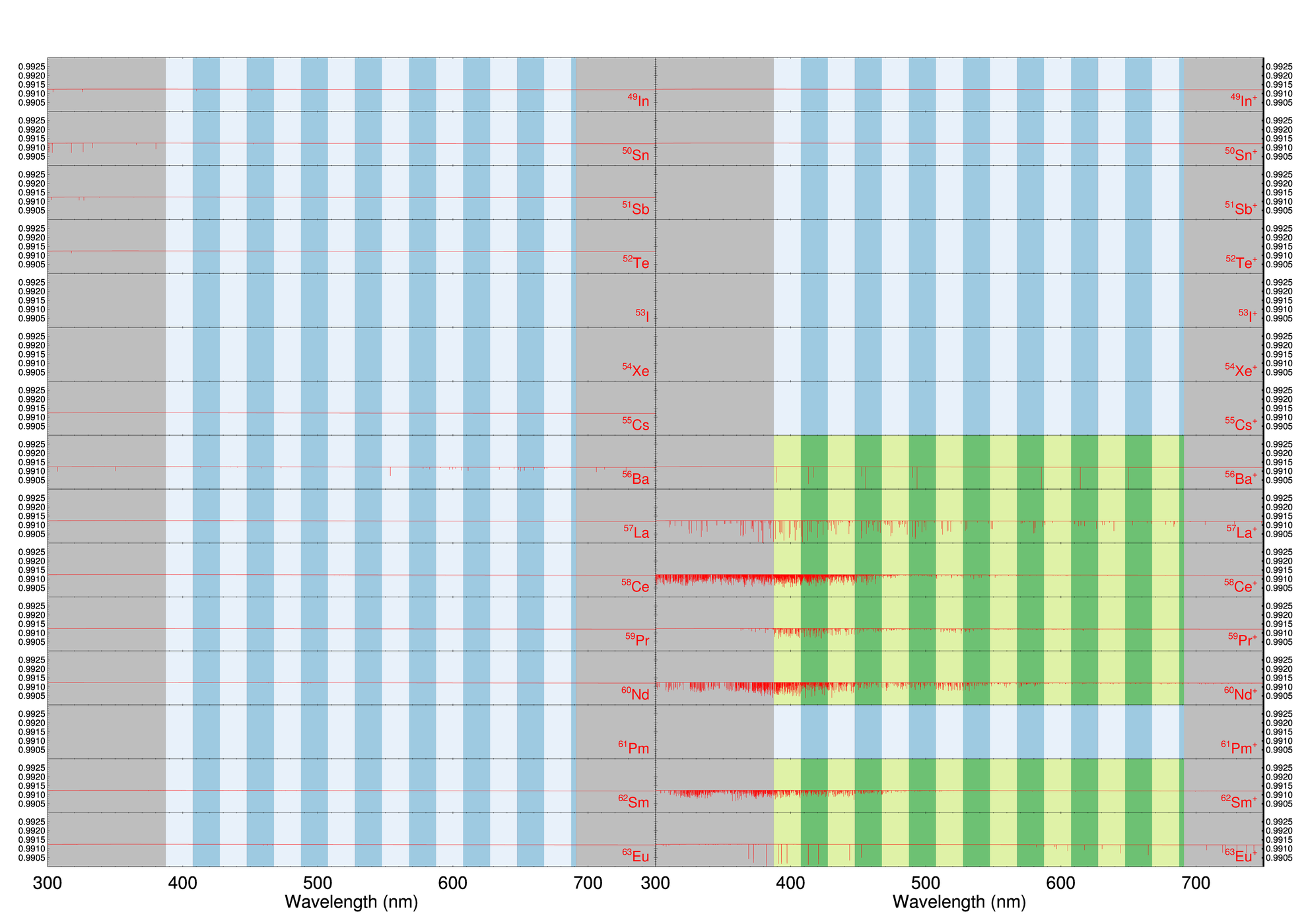}
\end{figure*}

\begin{figure*}
  \includegraphics[angle=90,width=18cm]{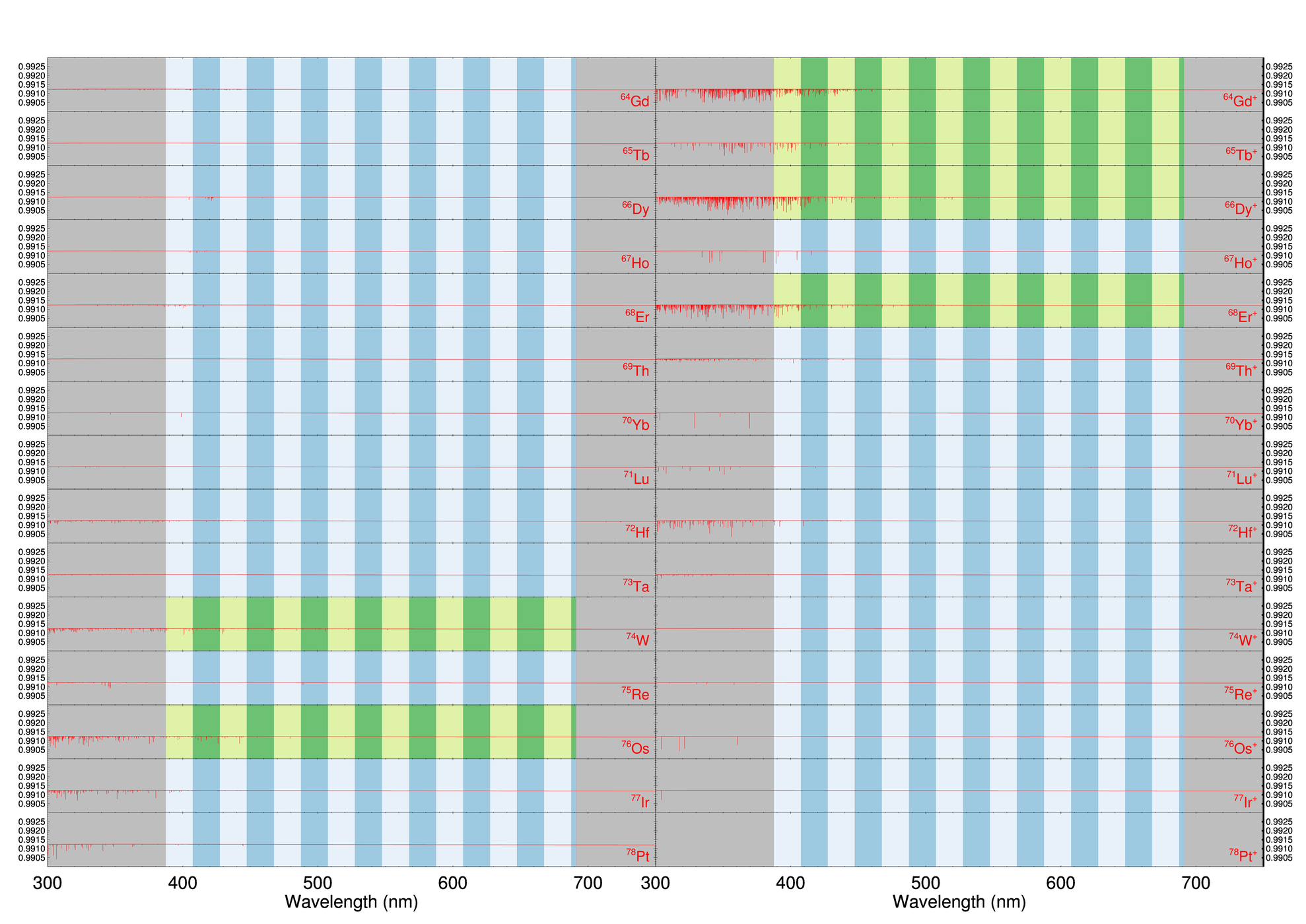}
\end{figure*}

\begin{figure*}
  \caption{Cross-correlation functions of all surveyed species highlighted in Fig. \ref{fig:spectra}. The colours indicate detections (green) and non-detections (red), as well as tentative signatures that occur at the expected radial and orbital velocities of the planet, but at low confidence and/or with an irregular structure (brown).}
  \smallskip
  \includegraphics[width=18cm]{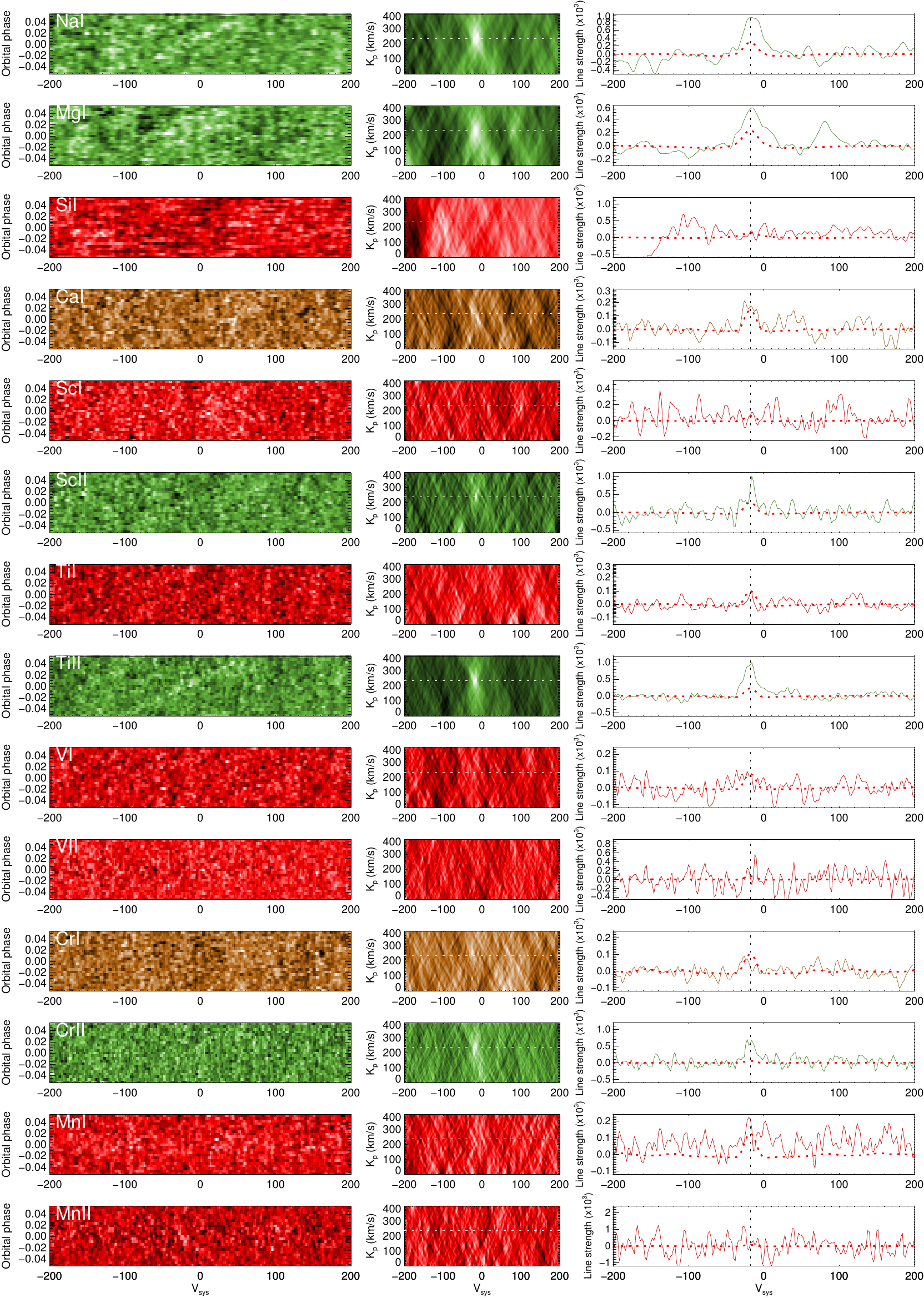}
  \label{fig:all_correlations}
\end{figure*}

\begin{figure*}
  \includegraphics[width=18cm]{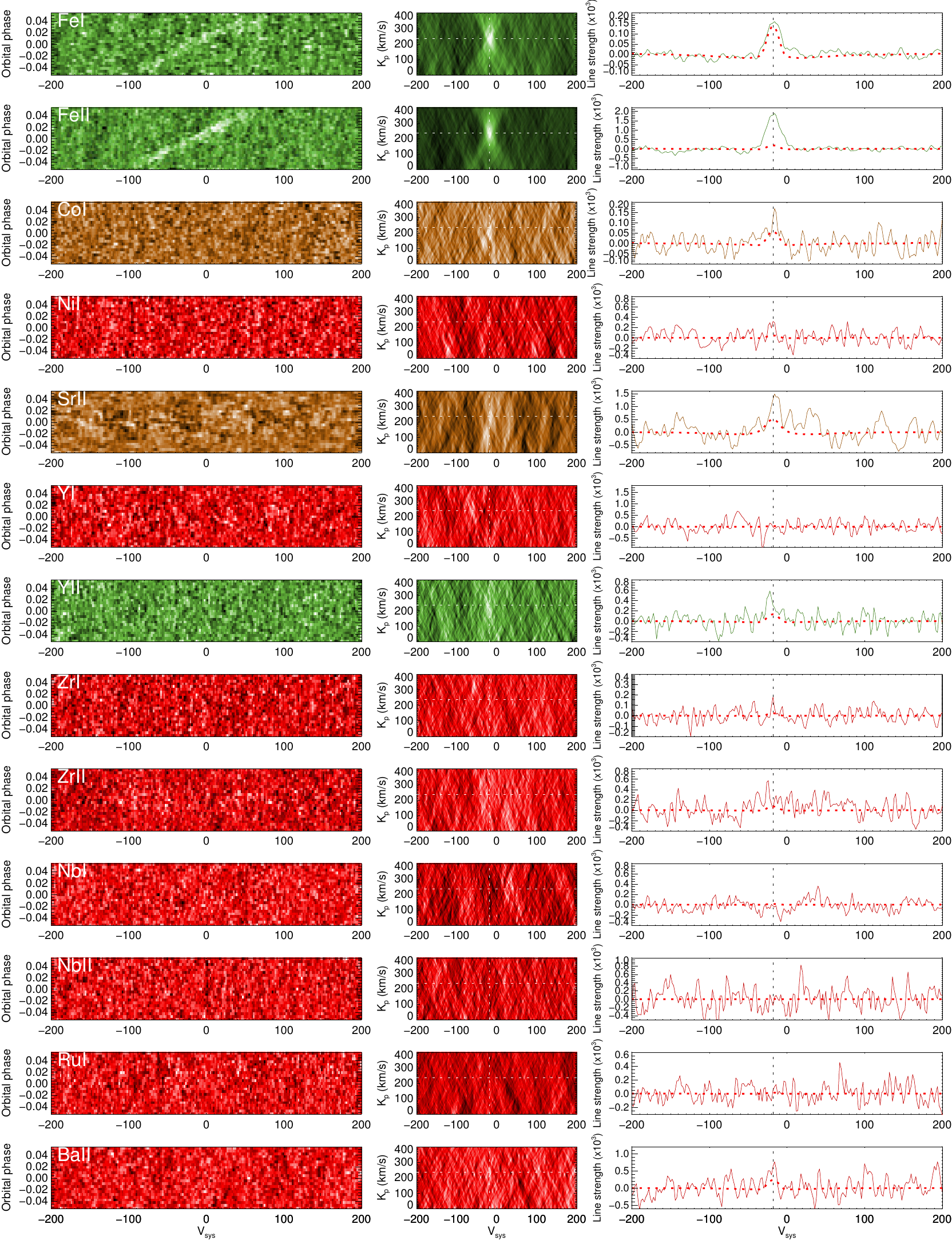}
\end{figure*}

\begin{figure*}
  \includegraphics[width=18cm]{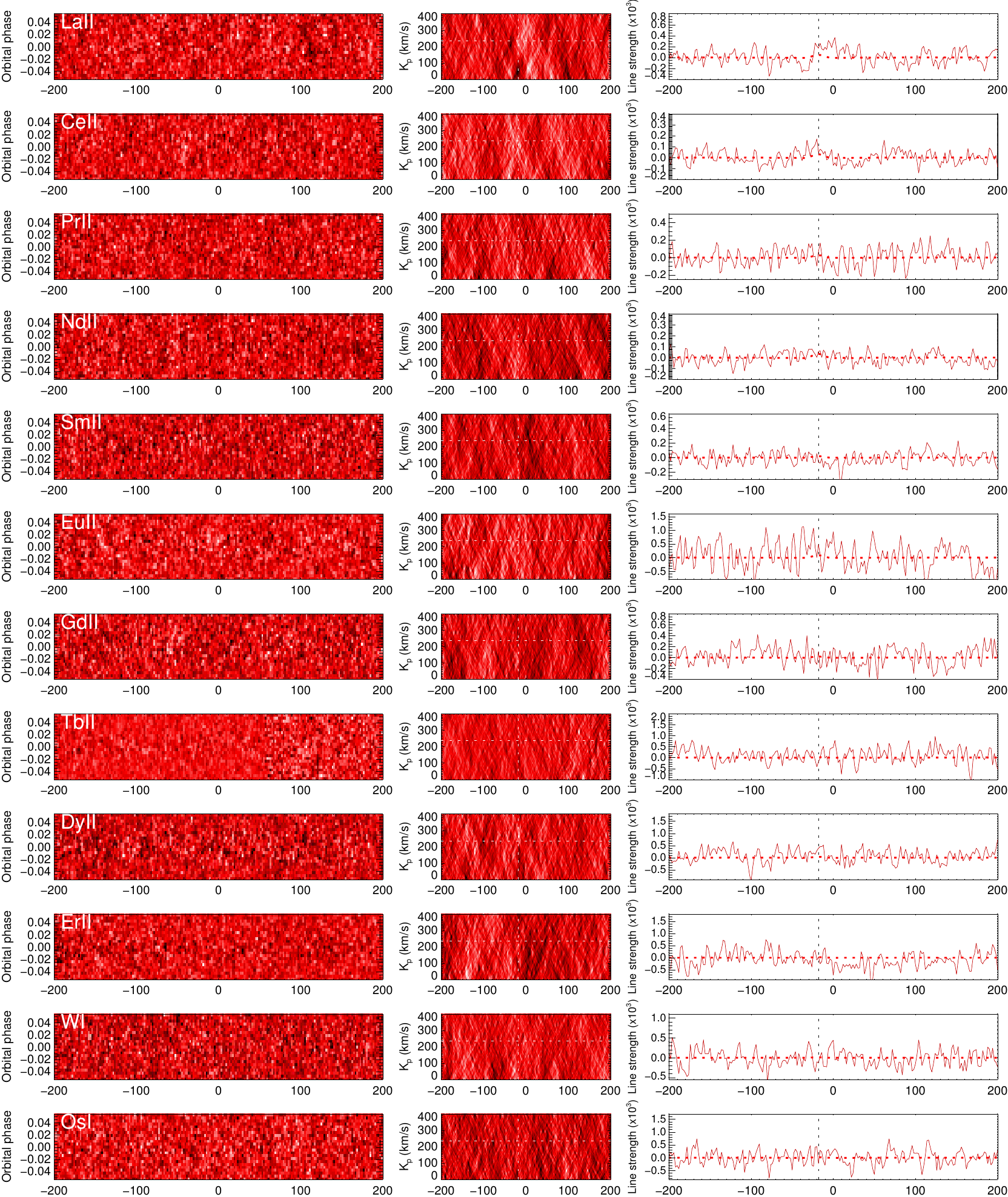}
\end{figure*}

\begin{table}
\caption{The scaling factors by which the model of the Doppler shadow was multiplied before subtracting it from the cross-correlation functions of each of the individual species, averaged over the two nights. These are reflective of the average depths of the absorption lines of these species in the stellar photosphere, times the average depth of the spectral lines in the PHOENIX photosphere model.}             
\label{table:DSfitting}      
\centering                          
\begin{tabular}{l c c}        
\hline\hline                 
Element		& Neutral	& Ion \\
\hline
\ion{Na}{}   &  0.786	& \\
\ion{Mg}{}   &  0.694 	& \\
\ion{Si}{}   &  0.130 	& \\
\ion{Ca}{}   &  0.161 	& \\
\ion{Sc}{}   &  0.0906 	&  0.497 \\
\ion{Ti}{}   &  0		&  0.704 \\
\ion{V }{}   &  0.00838  	&  0.110 \\
\ion{Cr}{}   &  0.0250      	&  0.573 \\
\ion{Mn}{}   &  0.0546      	&  0.106 \\
\ion{Fe}{}   &  0.110       	&  1.14 \\
\ion{Co}{}   &  0.0618      	& \\
\ion{Ni}{}   &  0.055		& \\
\ion{Sr}{}   &   		&  1.06 \\
\ion{Y }{}   &  0		&  0.189 \\
\ion{Zr}{}   &  0.000761 	&  0.0263 \\
\ion{Nb}{}   &  0		&  0.0675 \\
\ion{Ru}{}   &  0.00865	& \\
\ion{Ba}{}   &  		&  0.387 \\
\ion{La}{}   &  		&  0.102 \\
\ion{Ce}{}   &		&  0.00455 \\
\ion{Pr}{}   &		&  0.0291 \\
\ion{Nd}{}   &		&  0.00110 \\
\ion{Sm}{}   &		&  0 \\
\ion{Eu}{}   &		&  0.194 \\
\ion{Gd}{}   &		&  0.0266 \\
\ion{Tb}{}   &		&  0 \\
\ion{Dy}{}   &		&  0.107 \\
\ion{Ho}{}   &		&  0 \\
\ion{Er}{}   &		&  0.0902 \\
\ion{W }{}   &  0.142		& \\
\ion{Os}{}   &  0		& \\
\hline                                   
\end{tabular}
\end{table}

\clearpage
\onecolumn
\setcounter{table}{2}

   \begin{longtable}{l r r r r r l}
      \caption{\label{table:FastChemData} Coefficients for the fit of the temperature-dependent mass action constants of ions (including anions). The fits are valid in the temperature range between 100 K and 6000 K and are given in the \texttt{FastChem} format: $\ln \overline{K}_i(T) = \frac{a_0}{T} + a_1 \ln T + b_0 + b_1 T + b_2 T^2$ \citep{Stock2018}.}\\
            \hline\hline
            \noalign{\smallskip}
            Species & a$_0$ & a$_1$ & b$_0$ & b$_1$ & b$_2$ & Reference \\
            \noalign{\smallskip}
            \hline
            \endfirsthead
            \hline\hline
            \noalign{\smallskip}
            Species & a$_0$ & a$_1$ & b$_0$ & b$_1$ & b$_2$ & Reference \\
            \noalign{\smallskip}
            \hline
            \noalign{\smallskip}
            \endhead
            \hline
            \noalign{\smallskip}
            \endfoot
            \noalign{\smallskip}
            \ce{Li+} & -6.257651e+04 & 2.445125e+00 & -1.458903e+01 & 6.911880e-05 & -9.971750e-09 &  b \\
            \ce{Li^2+} & -9.403432e+05 & 4.954491e+00 & -2.817262e+01 & 5.864958e-05 & -8.710342e-09 &  b \\
            \ce{Li-} & 7.168737e+03 & -2.523718e+00 & 1.366656e+01 & 3.180655e-05 & -5.015196e-09 &  b \\
            \ce{Be+} & -1.081981e+05 & 2.448287e+00 & -1.320213e+01 & 4.864945e-05 & -5.720544e-09 &  a \\
            \ce{Be^2+} & -3.195220e+05 & 4.969213e+00 & -2.825779e+01 & 3.688612e-05 & -4.907970e-09 &  b \\
            \ce{B+} & -9.628053e+04 & 2.499047e+00 & -1.600591e+01 & 7.692046e-07 & -9.594323e-11 &  a \\
            \ce{B^2+} & -3.195039e+05 & 4.996323e+00 & -2.951576e+01 & 2.978475e-06 & -3.149719e-10 &  a \\
            \ce{B-} & 3.222835e+03 & -2.486416e+00 & 1.454006e+01 & -1.579851e-05 & 3.089472e-09 &  a \\
            \ce{C^2+} & -4.135903e+05 & 4.984436e+00 & -3.053864e+01 & 1.684809e-05 & -3.164118e-09 &  a \\
            \ce{N^2+} & -5.122099e+05 & 5.230785e+00 & -2.965536e+01 & -1.073986e-04 & 6.073966e-09 &  a \\
            \ce{O^2+} & -5.656110e+05 & 5.242922e+00 & -3.012987e+01 & -1.230956e-04 & 8.130114e-09 &  a \\
            \ce{F^2+} & -6.080289e+05 & 4.758626e+00 & -2.708094e+01 & 9.538401e-05 & -5.702377e-09 &  a \\
            \ce{Ne^2+} & -7.255567e+05 & 5.331451e+00 & -2.883738e+01 & -8.033248e-05 & 3.091043e-09 &  a \\
            \ce{Na^2+} & -6.083678e+05 & 4.994249e+00 & -2.778162e+01 & 1.569849e-04 & -1.785986e-08 &  a \\
            \ce{Mg^2+} & -2.632132e+05 & 4.967207e+00 & -2.824584e+01 & 3.926133e-05 & -5.215984e-09 &  a \\
            \ce{Al^2+} & -2.879089e+05 & 4.883613e+00 & -2.873766e+01 & 6.361646e-05 & -4.708568e-09 &  a \\
            \ce{Si^2+} & -2.841943e+05 & 4.782967e+00 & -2.910203e+01 & 1.008866e-04 & -1.017122e-08 &  a \\
            \ce{P^2+} & -3.510207e+05 & 5.702043e+00 & -3.321286e+01 & -2.354251e-04 & 6.150734e-09 &  a \\
            \ce{S^2+} & -3.910012e+05 & 5.789528e+00 & -3.409596e+01 & -3.487582e-04 & 2.160394e-08 &  a \\
            \ce{Cl^2+} & -4.268605e+05 & 4.843548e+00 & -2.745382e+01 & -1.815412e-05 & 5.086697e-09 &  a \\
            \ce{Ar^2+} & -5.034790e+05 & 5.165356e+00 & -2.791532e+01 & 5.757654e-05 & -6.291260e-09 &  a \\
            \ce{K^2+} & -4.174481e+05 & 4.800569e+00 & -2.660964e+01 & 3.187530e-04 & -3.631262e-08 &  a \\
            \ce{Ca^2+} & -2.087315e+05 & 4.826088e+00 & -2.741512e+01 & 2.233377e-04 & -3.281908e-08 &  a \\
            \ce{Ca-} & 2.491226e+02 & -2.330029e+00 & 1.530053e+01 & -6.796035e-05 & 5.088252e-09 &  a \\
            \ce{Sc+} & -7.630457e+04 & 1.939671e+00 & -1.029001e+01 & 5.553202e-04 & -4.992972e-08 &  c \\
            \ce{Sc^2+} & -2.246849e+05 & 4.929991e+00 & -2.810383e+01 & 1.530566e-04 & -2.843184e-08 &  a \\
            \ce{Sc-} & 2.178975e+03 & -2.360998e+00 & 1.380622e+01 & 2.356529e-05 & 5.089711e-10 &  a \\
            \ce{Ti^2+} & -2.367560e+05 & 5.084696e+00 & -2.902846e+01 & -1.698708e-05 & -1.619924e-08 &  a \\
            \ce{V^2+} & -2.484095e+05 & 5.040063e+00 & -2.863070e+01 & -1.905291e-04 & 8.702925e-09 &  a \\
            \ce{Cr^2+} & -2.698671e+05 & 5.771943e+00 & -3.258554e+01 & -3.510903e-04 & 5.252151e-09 &  a \\
            \ce{Mn^2+} & -2.677758e+05 & 4.913883e+00 & -2.793082e+01 & 1.069395e-04 & -1.499061e-08 &  a \\
            \ce{Fe^2+} & -2.795485e+05 & 5.019119e+00 & -2.859377e+01 & 1.347711e-05 & -1.155154e-08 &  a \\
            \ce{Co^2+} & -2.896757e+05 & 5.142486e+00 & -2.926769e+01 & -2.023968e-04 & 9.480268e-09 &  a \\
            \ce{Ni^2+} & -2.995070e+05 & 4.693818e+00 & -2.697957e+01 & 1.608813e-04 & -1.118022e-08 &  a \\
            \ce{Cu^2+} & -3.251517e+05 & 4.903999e+00 & -2.684981e+01 & 2.482547e-04 & -2.988537e-08 &  a \\
            \ce{Zn^2+} & -3.174840e+05 & 4.996642e+00 & -2.842085e+01 & 3.752581e-06 & -4.509092e-10 &  a \\
            \ce{Ga+} & -6.980530e+04 & 1.607740e+00 & -9.320869e+00 & 3.015434e-04 & -1.633885e-08 &  b \\
            \ce{Ga^2+} & -3.077803e+05 & 4.331296e+00 & -2.435889e+01 & 1.712459e-04 & -6.810332e-09 &  b \\
            \ce{Ga-} & 5.057817e+03 & -1.891447e+00 & 1.006145e+01 & -2.051303e-04 & 1.451311e-08 &  b \\
            \ce{Ge+} & -9.176634e+04 & 1.425826e+00 & -7.483645e+00 & 5.963214e-04 & -4.297209e-08 &  b \\
            \ce{Ge^2+} & -2.767541e+05 & 3.722745e+00 & -2.089299e+01 & 3.763743e-04 & -2.132323e-08 &  b \\
            \ce{Ge-} & 1.413645e+04 & -3.747478e+00 & 2.298821e+01 & 3.220351e-04 & -9.528489e-09 &  b \\
            \ce{As+} & -1.132523e+05 & 3.756197e+00 & -2.390343e+01 & -1.167552e-04 & -8.836227e-09 &  b \\
            \ce{As^2+} & -3.298491e+05 & 4.732759e+00 & -2.765064e+01 & 4.718120e-04 & -4.177260e-08 &  b \\
            \ce{As-} & 9.334082e+03 & -2.584351e+00 & 1.486495e+01 & 2.373549e-04 & -2.365011e-08 &  b \\
            \ce{Se+} & -1.131779e+05 & 2.554645e+00 & -1.466664e+01 & -2.114424e-04 & 1.839430e-08 &  b \\
            \ce{Se^2+} & -3.590378e+05 & 5.062661e+00 & -3.069360e+01 & 4.448638e-04 & -3.987156e-08 &  b \\
            \ce{Se-} & 2.343896e+04 & -2.524085e+00 & 1.420115e+01 & -8.135546e-05 & 6.462952e-09 &  b \\
            \ce{Rb+} & -4.854618e+04 & 2.197062e+00 & -1.301519e+01 & 2.942980e-04 & -3.670703e-08 &  b \\
            \ce{Rb^2+} & -3.651322e+05 & 4.808020e+00 & -2.660509e+01 & 2.284562e-04 & -2.926039e-08 &  b \\
            \ce{Rb-} & 5.642096e+03 & -2.471924e+00 & 1.337032e+01 & -4.960086e-05 & 9.617959e-09 &  b \\
            \ce{Sr+} & -6.614589e+04 & 2.245360e+00 & -1.194103e+01 & 2.583965e-04 & -3.380599e-08 &  a \\
            \ce{Sr^2+} & -1.941161e+05 & 4.792835e+00 & -2.722238e+01 & 2.716941e-04 & -4.104279e-08 &  b \\
            \ce{Sr-} & 6.627934e+02 & -1.992876e+00 & 1.269203e+01 & -1.370573e-04 & 4.628944e-09 &  a \\
            \ce{Y+} & -7.176276e+04 & 4.188361e+00 & -2.641749e+01 & -4.324053e-04 & 7.814626e-09 &  a \\
            \ce{Y^2+} & -2.141923e+05 & 4.843393e+00 & -2.767314e+01 & 2.539009e-04 & -3.793062e-08 &  b \\
            \ce{Y-} & 3.556755e+03 & -2.634410e+00 & 1.507707e+01 & 4.561236e-04 & -3.724981e-08 &  a \\
            \ce{Zr+} & -7.701333e+04 & 2.627948e+00 & -1.484642e+01 & 3.778927e-05 & -1.404338e-08 &  a \\
            \ce{Zr^2+} & -2.293259e+05 & 5.104568e+00 & -2.914197e+01 & -1.287425e-04 & -5.281367e-09 &  a \\
            \ce{Zr-} & 5.029101e+03 & -2.350380e+00 & 1.428294e+01 & -1.913007e-04 & 8.780821e-09 &  a \\
            \ce{Nb+} & -7.857835e+04 & 2.612763e+00 & -1.580872e+01 & 1.123904e-04 & -1.049845e-08 &  a \\
            \ce{Nb^2+} & -2.445578e+05 & 5.211120e+00 & -3.088235e+01 & -3.620478e-05 & -5.447194e-09 &  a \\
            \ce{Nb-} & 1.058498e+04 & -3.066374e+00 & 1.704201e+01 & -8.271939e-05 & 2.007041e-08 &  a \\
            \ce{Mo+} & -8.230749e+04 & 2.489498e+00 & -1.430508e+01 & 6.234623e-06 & -4.526990e-10 &  a \\
            \ce{Mo^2+} & -2.697311e+05 & 6.302578e+00 & -3.729495e+01 & -2.766927e-04 & -3.447993e-09 &  a \\
            \ce{Mo-} & 8.620377e+03 & -2.969326e+00 & 1.665969e+01 & 8.507653e-04 & -7.317276e-08 &  a \\
            \ce{Ru+} & -8.543378e+04 & 2.412469e+00 & -1.375638e+01 & 6.152285e-07 & -3.017937e-09 &  a \\
            \ce{Ru^2+} & -2.798820e+05 & 5.157276e+00 & -2.959099e+01 & -1.427370e-04 & 6.698495e-11 &  a \\
            \ce{Ru-} & 1.211661e+04 & -2.677755e+00 & 1.518654e+01 & 1.408877e-04 & -1.581984e-08 &  a \\
            \ce{Rh+} & -8.660251e+04 & 2.382664e+00 & -1.346571e+01 & -9.551840e-05 & 7.588277e-09 &  a \\
            \ce{Rh^2+} & -2.963766e+05 & 4.971467e+00 & -2.820594e+01 & -1.055777e-04 & 7.448432e-09 &  a \\
            \ce{Rh-} & 1.317054e+04 & -2.981142e+00 & 1.499211e+01 & 1.587576e-04 & 1.303818e-08 &  a \\
            \ce{Pd+} & -9.650404e+04 & 3.427600e+00 & -1.837233e+01 & -6.848823e-04 & 2.286147e-08 &  a \\
            \ce{Pd^2+} & -3.221492e+05 & 5.425302e+00 & -2.885909e+01 & -3.406978e-04 & 1.232565e-09 &  a \\
            \ce{Pd-} & 6.619238e+03 & -1.901975e+00 & 1.129693e+01 & -5.718141e-04 & 1.219657e-08 &  a \\
            \ce{Ag+} & -8.791960e+04 & 2.494306e+00 & -1.487763e+01 & 5.114261e-06 & -5.635751e-10 &  b \\
            \ce{Ag^2+} & -3.373123e+05 & 4.913882e+00 & -2.683080e+01 & 9.621756e-05 & -3.857636e-09 &  b \\
            \ce{Ag-} & 1.513778e+04 & -2.499833e+00 & 1.352636e+01 & -1.864851e-07 & 2.239051e-11 &  b \\
            \ce{Cd+} & -1.043706e+05 & 2.492493e+00 & -1.347983e+01 & 6.696760e-06 & -7.311102e-10 &  b \\
            \ce{Cd^2+} & -3.005824e+05 & 4.995789e+00 & -2.841579e+01 & 4.798662e-06 & -5.945109e-10 &  b \\
            \ce{In+} & -6.720764e+04 & 2.430559e+00 & -1.421171e+01 & -2.867153e-04 & 2.412735e-08 &  b \\
            \ce{In^2+} & -2.861252e+05 & 5.086195e+00 & -2.879086e+01 & -3.773571e-04 & 3.074917e-08 &  b \\
            \ce{In-} & 4.494223e+03 & -2.363480e+00 & 1.250462e+01 & 2.701248e-04 & -2.307383e-08 &  b \\
            \ce{Sn+} & -8.550616e+04 & 1.636183e+00 & -7.583424e+00 & -9.394052e-06 & 1.328952e-08 &  b \\
            \ce{Sn^2+} & -2.550640e+05 & 4.949683e+00 & -2.776817e+01 & -6.474905e-04 & 5.382399e-08 &  b \\
            \ce{Sn-} & 1.286251e+04 & -2.549000e+00 & 1.628734e+01 & -6.778121e-04 & 6.572371e-08 &  b \\
            \ce{Sb+} & -9.999992e+04 & 1.861317e+00 & -1.187724e+01 & 8.992890e-04 & -6.835289e-08 &  b \\
            \ce{Sb^2+} & -2.917412e+05 & 4.893050e+00 & -2.849750e+01 & 1.207286e-04 & -9.306244e-09 &  b \\
            \ce{Sb-} & 1.214207e+04 & -2.590659e+00 & 1.497120e+01 & 1.213584e-04 & -1.205327e-08 &  b \\
            \ce{Te+} & -1.048358e+05 & 1.636178e+00 & -7.583391e+00 & -9.390886e-06 & 1.328928e-08 &  b \\
            \ce{Te^2+} & -3.204477e+05 & 4.659802e+00 & -2.802328e+01 & 3.713875e-04 & -1.538476e-08 &  b \\
            \ce{Te-} & 2.288286e+04 & -2.421080e+00 & 1.352850e+01 & -8.725159e-05 & 2.439641e-09 &  b \\
            \ce{Cs+} & -4.528316e+04 & 2.087691e+00 & -1.234395e+01 & 4.185307e-04 & -5.552924e-08 &  b \\
            \ce{Cs^2+} & -3.139541e+05 & 4.755044e+00 & -2.630289e+01 & 3.156288e-04 & -4.718043e-08 &  b \\
            \ce{Cs-} & 5.399787e+03 & -2.931135e+00 & 1.617382e+01 & 3.635188e-04 & -4.840012e-09 &  b \\
            \ce{Ba+} & -6.051156e+04 & 2.230065e+00 & -1.201430e+01 & 4.601799e-04 & -6.798839e-08 &  b \\
            \ce{Ba^2+} & -1.765336e+05 & 5.179498e+00 & -2.960346e+01 & -3.866059e-05 & -4.383653e-08 &  b \\
            \ce{Ba-} & 1.750082e+03 & -2.233250e+00 & 1.365346e+01 & 4.925290e-04 & -7.194586e-08 &  b \\
            \ce{La+} & -6.451483e+04 & 3.326088e+00 & -1.927360e+01 & -4.769959e-04 & 1.819639e-08 &  b \\
            \ce{La^2+} & -1.930757e+05 & 4.975852e+00 & -2.818165e+01 & -2.888187e-04 & 1.285448e-08 &  b \\
            \ce{La-} & 6.348587e+03 & -2.654876e+00 & 1.605691e+01 & 4.380221e-04 & -3.161517e-08 &  b \\
            \ce{Ce+} & -6.403419e+04 & 3.171264e+00 & -1.914062e+01 & -1.056822e-04 & -1.019987e-08 &  b \\
            \ce{Ce^2+} & -1.902176e+05 & 4.699222e+00 & -2.673118e+01 & -1.068922e-04 & 1.316049e-09 &  b \\
            \ce{Ce-} & 6.548959e+03 & -2.955035e+00 & 1.702923e+01 & -3.701637e-05 & 2.383397e-08 &  b \\
            \ce{Pr+} & -6.342175e+04 & 3.126189e+00 & -1.798169e+01 & -2.761204e-04 & 3.578332e-09 &  b \\
            \ce{Pr^2+} & -1.858490e+05 & 5.582066e+00 & -3.192867e+01 & -6.136215e-04 & 2.352541e-08 &  b \\
            \ce{Pr-} & 1.122440e+04 & -2.082943e+00 & 1.164134e+01 & -4.125150e-04 & 3.774982e-08 &  b \\
            \ce{Nd+} & -6.403886e+04 & 2.975283e+00 & -1.718373e+01 & -7.720646e-05 & -1.314762e-08 &  b \\
            \ce{Nd^2+} & -1.885906e+05 & 5.217612e+00 & -2.981865e+01 & -1.050506e-04 & -3.399976e-08 &  b \\
            \ce{Nd-} & 2.230139e+04 & -2.102172e+00 & 1.134207e+01 & -7.139634e-05 & -3.701865e-08 &  b \\
            \ce{Sm+} & -6.547008e+04 & 2.282160e+00 & -1.268932e+01 & 3.636024e-04 & -3.906499e-08 &  b \\
            \ce{Sm^2+} & -1.939742e+05 & 4.842841e+00 & -2.753764e+01 & 2.379702e-04 & -4.274995e-08 &  b \\
            \ce{Sm-} & 1.758414e+03 & -3.749298e+00 & 2.321110e+01 & 5.770456e-05 & 7.703514e-09 &  b \\
            \ce{Eu+} & -6.581677e+04 & 2.345689e+00 & -1.328486e+01 & 3.752787e-04 & -4.408329e-08 &  b \\
            \ce{Eu^2+} & -1.962769e+05 & 4.799913e+00 & -2.727418e+01 & 2.795278e-04 & -4.568226e-08 &  b \\
            \ce{Eu-} & 1.388057e+03 & -1.804402e+00 & 1.017602e+01 & -7.181045e-06 & -5.833591e-09 &  b \\
            \ce{Gd+} & -7.137337e+04 & 2.276964e+00 & -1.300354e+01 & 4.336840e-04 & -4.546894e-08 &  b \\
            \ce{Gd^2+} & -2.116502e+05 & 4.932654e+00 & -2.835878e+01 & 2.831846e-04 & -4.332661e-08 &  b \\
            \ce{Gd-} & 1.546839e+03 & -2.916868e+00 & 1.763578e+01 & 3.829147e-04 & -3.072644e-08 &  b \\
            \ce{Tb+} & -6.799838e+04 & 2.010817e+00 & -1.204494e+01 & 5.350978e-04 & -4.895898e-08 &  b \\
            \ce{Tb^2+} & -2.018045e+05 & 4.200660e+00 & -2.406078e+01 & 2.802467e-04 & -1.396694e-08 &  b \\
            \ce{Tb-} & 1.348067e+04 & -3.028296e+00 & 1.710919e+01 & -1.334802e-04 & 3.541371e-08 &  b \\
            \ce{Dy+} & -6.879472e+04 & 3.014676e+00 & -1.747181e+01 & -1.157870e-04 & -1.117657e-08 &  b \\
            \ce{Dy^2+} & -2.043326e+05 & 5.041662e+00 & -2.873304e+01 & 4.613373e-05 & -4.562549e-08 &  b \\
            \ce{Dy-} & 4.084086e+03 & -2.493290e+00 & 1.413667e+01 & -2.962651e-05 & 7.644436e-10 &  b \\
            \ce{Ho+} & -6.977167e+04 & 2.996212e+00 & -1.728031e+01 & -8.020657e-05 & -2.656349e-08 &  b \\
            \ce{Ho^2+} & -2.068032e+05 & 5.015257e+00 & -2.857871e+01 & 6.703620e-05 & -3.148431e-08 &  b \\
            \ce{Ho-} & 3.906039e+03 & -2.604632e+00 & 1.464065e+01 & 1.087372e-04 & -5.276441e-09 &  b \\
            \ce{Er+} & -7.078944e+04 & 3.047784e+00 & -1.739668e+01 & -2.387930e-04 & -1.836723e-09 &  b \\
            \ce{Er^2+} & -2.093003e+05 & 5.089075e+00 & -2.902460e+01 & -3.946951e-06 & -3.000136e-08 &  b \\
            \ce{Er-} & 3.682983e+03 & -2.119010e+00 & 1.141914e+01 & -3.518785e-04 & 8.140888e-09 &  b \\
            \ce{Tm+} & -7.184576e+04 & 2.426255e+00 & -1.322566e+01 & 1.704652e-04 & -2.721242e-08 &  b \\
            \ce{Tm^2+} & -2.116323e+05 & 4.782511e+00 & -2.715946e+01 & 2.813775e-04 & -4.166156e-08 &  b \\
            \ce{Tm-} & 1.193280e+04 & -2.565770e+00 & 1.251438e+01 & 1.095083e-04 & -2.155381e-08 &  b \\
            \ce{Yb+} & -7.262584e+04 & 2.296951e+00 & -1.225092e+01 & 1.917634e-04 & -2.271257e-08 &  b \\
            \ce{Yb^2+} & -2.138997e+05 & 4.833709e+00 & -2.745114e+01 & 1.989037e-04 & -2.642820e-08 &  b \\
            \ce{Lu+} & -6.300162e+04 & 2.540992e+00 & -1.557599e+01 & -4.279115e-04 & 5.199291e-08 &  b \\
            \ce{Lu^2+} & -2.243108e+05 & 4.820128e+00 & -2.789396e+01 & -9.521387e-05 & 1.856125e-08 &  b \\
            \ce{Lu-} & 3.973630e+03 & -2.756844e+00 & 1.600575e+01 & 2.107913e-04 & -5.906525e-09 &  b \\
            \ce{Hf+} & -7.928592e+04 & 2.161891e+00 & -1.229210e+01 & 2.526805e-04 & -1.769312e-08 &  b \\
            \ce{Hf^2+} & -2.520756e+05 & 5.292591e+00 & -3.007134e+01 & -5.024157e-04 & 2.303820e-08 &  b \\
            \ce{Hf-} & 2.062321e+03 & -2.547061e+00 & 1.423421e+01 & 1.360278e-04 & -5.484156e-09 &  b \\
            \ce{Ta+} & -8.744604e+04 & 3.158873e+00 & -1.877288e+01 & -1.554275e-04 & -1.253218e-09 &  b \\
            \ce{Ta^2+} & -2.755660e+05 & 5.339651e+00 & -3.030819e+01 & -6.384433e-04 & 2.852234e-08 &  b \\
            \ce{Ta-} & 3.847936e+03 & -2.093770e+00 & 9.994987e+00 & 3.261281e-04 & -4.536083e-08 &  b \\
            \ce{W+} & -9.150081e+04 & 1.776468e+00 & -8.550768e+00 & 8.439088e-05 & 7.391207e-09 &  b \\
            \ce{W^2+} & -2.813461e+05 & 4.382890e+00 & -2.444833e+01 & 7.191376e-05 & 1.445050e-08 &  b \\
            \ce{W-} & 9.432156e+03 & -2.405402e+00 & 1.604370e+01 & -1.189007e-03 & 1.106563e-07 &  b \\
            \ce{Re+} & -9.091879e+04 & 2.421065e+00 & -1.359550e+01 & 1.070258e-04 & -1.897499e-08 &  b \\
            \ce{Re^2+} & -2.835463e+05 & 4.945248e+00 & -2.812904e+01 & 8.913961e-05 & -1.714456e-08 &  b \\
            \ce{Re-} & 6.391571e+02 & -2.938058e+00 & 1.718643e+01 & 5.689440e-04 & -4.466452e-08 &  b \\
            \ce{Os+} & -9.790556e+04 & 2.578877e+00 & -1.459216e+01 & -9.155119e-05 & 2.202461e-09 &  b \\
            \ce{Os^2+} & -2.951961e+05 & 5.019104e+00 & -2.854965e+01 & -2.882173e-05 & -4.096798e-09 &  b \\
            \ce{Os-} & 1.251071e+04 & -2.487400e+00 & 1.423524e+01 & 1.393521e-05 & -5.261997e-09 &  b \\
            \ce{Ir+} & -1.040089e+05 & 2.669729e+00 & -1.524435e+01 & -7.712863e-05 & 2.059896e-09 &  b \\
            \ce{Ir^2+} & -3.012993e+05 & 5.264280e+00 & -2.997023e+01 & -3.584437e-04 & 1.315339e-08 &  b \\
            \ce{Ir-} & 1.829105e+04 & -1.568548e+00 & 8.278251e+00 & -5.396383e-04 & 2.501637e-08 &  b \\
            \ce{Pt+} & -1.042118e+05 & 1.320760e+00 & -7.001288e+00 & 5.594756e-04 & -2.525852e-08 &  b \\
            \ce{Pt^2+} & -3.194780e+05 & 4.332828e+00 & -2.413334e+01 & 1.811947e-04 & -1.186298e-08 &  b \\
            \ce{Pt-} & 2.455609e+04 & -3.174591e+00 & 1.708680e+01 & 1.556494e-04 & 1.006783e-09 &  b \\
            \ce{Au+} & -1.070045e+05 & 2.707671e+00 & -1.623589e+01 & -1.701164e-04 & 1.276849e-08 &  b \\
            \ce{Au^2+} & -3.449473e+05 & 5.007420e+00 & -2.740510e+01 & 2.554497e-05 & -1.286974e-08 &  b \\
            \ce{Au-} & 2.679326e+04 & -2.492163e+00 & 1.346216e+01 & 2.508001e-05 & -1.281389e-08 &  b \\
            \ce{Hg+} & -1.211222e+05 & 2.500164e+00 & -1.352840e+01 & -1.513729e-07 & 1.719679e-11 &  b \\
            \ce{Hg^2+} & -3.387766e+05 & 4.999857e+00 & -2.844015e+01 & 1.615884e-07 & -1.969555e-11 &  b \\
            \ce{Tl+} & -7.085577e+04 & 2.594299e+00 & -1.553214e+01 & -4.979836e-05 & -5.224852e-09 &  b \\
            \ce{Tl^2+} & -3.079320e+05 & 5.042748e+00 & -2.871117e+01 & -1.973543e-05 & -7.424893e-09 &  b \\
            \ce{Tl-} & 4.366684e+03 & -2.538768e+00 & 1.378263e+01 & 5.992629e-07 & 1.267056e-08 &  b \\
            \ce{Pb+} & -8.599945e+04 & 2.735546e+00 & -1.507083e+01 & -1.299813e-04 & -9.032662e-09 &  b \\
            \ce{Pb^2+} & -2.604921e+05 & 5.081516e+00 & -2.896644e+01 & -2.033514e-05 & -2.009540e-08 &  b \\
            \ce{Pb-} & 4.160111e+03 & -2.388659e+00 & 1.490846e+01 & -6.395885e-05 & -1.246717e-08 &  b \\
            \ce{Bi+} & -8.452993e+04 & 2.569671e+00 & -1.603952e+01 & -7.216128e-05 & 9.600723e-09 &  b \\
            \ce{Bi^2+} & -2.782281e+05 & 4.979024e+00 & -2.901385e+01 & 3.292111e-05 & -6.184604e-09 &  b \\
            \ce{Bi-} & 1.093209e+04 & -2.525123e+00 & 1.459062e+01 & 3.398221e-05 & -5.099485e-09 &  b \\
            \ce{Th+} & -7.304474e+04 & 2.979942e+00 & -1.767323e+01 & -2.152028e-05 & -1.087038e-08 &  b \\
            \ce{Th^2+} & -2.066111e+05 & 5.368242e+00 & -2.958683e+01 & -2.757084e-04 & -1.065319e-09 &  b \\
            \ce{U+} & -7.187899e+04 & 2.500000e+00 & -1.448287e+01 & -1.221136e-11 & 9.189032e-16 &  b \\
            \ce{U^2+} & -2.064914e+05 & 5.000000e+00 & -2.880874e+01 & -3.535195e-15 & 2.891645e-19 &  b \\
   \end{longtable}
   \tablefoot{\textbf{a}: Partition functions based on data from the Kurucz line lists \citep{Kurucz1995all..book.....K}. \textbf{b}: Partition functions based on data from the NIST Atomic Spectra Database \citep{NIST_ASD}. The ionisation energies are taken from the \citet{CRC_handbook85Ed} in both cases. \textbf{c}: Based on thermochemical data from \citet{Gurvich1982}.}

\end{appendix}


\bibliographystyle{bibtex/aa.bst}
\bibliography{bib.bib}

\end{document}